\documentclass[aps,prd,twocolumn,nofootinbib,longbibliography,superscriptaddress,aip]{revtex4-1}
\usepackage{hyperref}
\usepackage{amsmath}
\usepackage{amssymb}
\usepackage{amsthm}
\usepackage{graphicx}
\usepackage{physics}
\usepackage{xcolor}
\usepackage{mathrsfs}
\usepackage{dsfont}
\def\be{\begin{equation}}
\def\ee{\end{equation}}
\definecolor{bluenice}{rgb}{0.0, 0.53, 0.74}
\definecolor{princetonorange}{rgb}{1.0, 0.56, 0.0}

\newcommand{\Sbh}{S}
\newcommand{\rd}{\mathrm{d}}

\begin{document}

\title{Universal Construction of Black Hole Microstates}

\author{Ana Climent}
\affiliation{Departament de F\'isica Qu\`antica i Astrof\'isica, Institut de Ci\`encies del Cosmos, Universitat de Barcelona, Mart\'i i Franqu\`es 1, E-08028 Barcelona, Spain}
\author{Roberto Emparan}
\affiliation{Departament de F\'isica Qu\`antica i Astrof\'isica, Institut de Ci\`encies del Cosmos, Universitat de Barcelona, Mart\'i i Franqu\`es 1, E-08028 Barcelona, Spain}
\affiliation{Institució Catalana de Recerca i Estudis Avançats (ICREA), Passeig Lluis Companys, 23, 08010 Barcelona, Spain}
\author{Javier M.~Mag\'{a}n}
\affiliation{Instituto Balseiro, Centro At\'omico Bariloche, 8400-S.C. de Bariloche, R\'io Negro, Argentina}
\author{Martin~Sasieta}
\affiliation{Martin Fisher School of Physics, Brandeis University, Waltham, Massachusetts 02453, USA}
\author{Alejandro Vilar L\'opez}
\affiliation{Physique Théorique et Mathématique and International Solvay Institutes, Université Libre de Bruxelles (ULB), C.P. 231, 1050 Brussels, Belgium}

\begin{abstract}
We refine and extend a recent construction of sets of black hole microstates with semiclassical interiors that span a Hilbert space of dimension $e^S$, where $S$ is the black hole entropy. We elaborate on the definition and properties of microstates in statistical and black hole mechanics. The gravitational description of microstates employs matter shells in the interior of the black hole, and we argue that in the limit where the shells are very heavy, the construction acquires universal validity. To this end, we show it for very wide classes of black holes: we first extend the construction to rotating and charged black holes, including extremal and near-extremal solutions, with or without supersymmetry, and we sketch how the construction of microstates can be embedded in String Theory. We then describe how the approach can include general quantum corrections, near or far from extremality. 
For supersymmetric black holes, the microstates we construct differ from other recent constructions in that the interior excitations are not confined within the near-extremal throat. 
\end{abstract}

\maketitle

\section{Introduction} 

All the existing derivations of the black hole entropy formula \cite{Bekenstein:1973ur,Hawking:1975vcx}
\begin{equation}\label{bhe} 
S=\frac{A}{4G\hbar}
\end{equation}
have something unsatisfactory to them, and this dissatisfaction has stimulated much investigation into the structure of spacetime. In this article, following and extending the work of \cite{Balasubramanian:2022gmo,Balasubramanian:2022lnw}, we will develop a very broad statistical interpretation of this formula. To set the stage, we begin with a lightning review of \eqref{bhe} and its discontents.

\paragraph*{Not counting states.} The initial phenomenological arguments of Bekenstein and Hawking \cite{Bekenstein:1973ur,Hawking:1975vcx} hardly gave any hint about the statistical origin of this entropy. But soon after, Gibbons and Hawking offered a seemingly more fundamental derivation \cite{Gibbons:1976ue}. They interpreted the Euclidean gravitational path integral (GPI) as computing the (grand-)canonical partition function of gravity, from which one readily recovers \eqref{bhe} for large black holes. This success is as remarkable as it is puzzling: 
there are no microstates in sight, and therefore no interpretation as a state counting in quantum statistical mechanics. Technically, the entropy arises from a classical saddle point value of the integral, and not as a one-loop trace in a Hilbert space. Moreover, the result is universal, i.e., valid for any theory of Einstein-Hilbert gravity, regardless of its matter content and whether there exists, or not, a consistent quantum theory behind it. 

\paragraph*{Counting states but not black holes.} String Theory has provided detailed microscopic accounts of this entropy \cite{Strominger:1996sh}. However, what is counted here are the states of a weakly gravitating system that is not a black hole. 
To make plain the nature of actual black hole microstates, it seems necessary to remain in the strong gravity regime, and preferably without supersymmetry.

\paragraph*{Counting black hole states, with wormhole statistics.} A novel use of the GPI has made significant headway in identifying genuinely gravitational black hole microstates, indeed complete sets that comprise a Hilbert space of dimension $e^{S}$. This application of the GPI is consistent with \cite{Gibbons:1976ue}, but more fundamental than it in that it provides a statistical-mechanical interpretation. The microstates are effectively realized as smooth geometries with the same exterior as the black hole, but with different interiors. The GPI is employed to prepare a large number of quantum states, and to compute the overlaps between them to obtain the dimension of the space they span. On the face of it, this is a conventional use of the Euclidean path integral for a quantum field theory. However, to succeed in rendering the correct finite dimension of the Hilbert space, the GPI (a low-energy effective tool, oblivious to ultraviolet structure) resorts to a perplexing maneuver: Euclidean wormholes are necessary, and they confer an intrinsically statistical character to the overlaps between gravitational microstates. This approach to the microstates of black holes and their counting has been given its most general form so far in \cite{Balasubramanian:2022gmo,Balasubramanian:2022lnw}, building up on previous work on geometric microstates in two-dimensional theories \cite{Kourkoulou:2017zaj,Goel:2018ubv,Penington:2019kki,Hsin:2020mfa,Lin:2022rzw,Lin:2022zxd} and in higher-dimensions \cite{Chandra:2022fwi,Chandra:2023rhx}. More recently it has been used in two-dimensional supersymmetric scenarios \cite{Boruch:2023trc} and cosmological spacetimes \cite{Antonini:2023hdh}.

Our main goal is to provide strong evidence for the universal applicability of these constructions---a universality as wide as that of the Gibbons-Hawking derivation of \eqref{bhe}, but stemming from a different and more intricate analysis. 
We will do this along several fronts. The first and most natural one is to include black holes with rotation and charge. Afterwards, we are led to analyze the extension to supersymmetric black holes, which is important in the attempt to embed the construction within String Theory and to connect to microscopic string descriptions. Finally, we will ascertain how this construction accommodates quantum loop corrections to black hole entropy---the familiar $\log A$ terms, but also the more important $\log T$ effects near extremality that have recently attracted attention \cite{Maldacena:2016upp, Stanford:2017thb,Ghosh:2019rcj,Iliesiu:2020qvm,Heydeman:2020hhw,Boruch:2022tno,Iliesiu:2022onk,Banerjee:2023quv,Kapec:2023ruw,Rakic:2023vhv,Banerjee:2023gll,Emparan:2023ypa}. 

The construction of these microstates relies on the use of matter shells in the interior of the black hole. We will see that its large degree of universality is most transparent in the limit where the shells are very heavy. In this limit, the shells create large interiors and become asymptotically independent of any properties of the black hole other than its mass, charge and spin. They also become effectively independent of one another.
Our arguments thus require arbitrarily heavy shells, but this is not problematic: the shells lie entirely within the black hole interior, so they are screened and the exterior geometry is that of a black hole of fixed, finite macroscopic parameters.

\medskip

The plan for the remainder of the article is as follows: Section~\ref{sec:countingstates} introduces the basic strategy to count the dimension of the Hilbert space of a set of states with access to the moments of their mutual overlaps. Section~\ref{sec:BHmicrostates} discusses the general definition and properties of black hole microstates and their construction as Partially Entangled Grand-Canonical States (PEGS). We give explicit details for microstates with charge and rotation and their embedding in String Theory. Section~\ref{sec:countingmicro} applies the strategy to the computation of the dimension of the Hilbert space of black hole microstates. We describe two different methods to derive the microcanonical moments of the matrix of overlaps between microstates. We show that, in the appropriate (heavy-shell) limit, both methods lead to a simple, universal result for these moments, which implies the universality of the derivation of the black hole entropy.  Section~\ref{sec:corrections} presents arguments to the effect that the same construction can easily account for quantum corrections (near extremality and more generally) and statistical corrections to the black hole density of states. We conclude in Section~\ref{sec:discussion} summarizing the implications of our study and discussing open questions. Appendix \hyperref[app:FreeProbability]{A} is a more general analysis of junction conditions in Eintein-Maxwell theories, in particular allowing surface currents on the shell. Appendix \hyperref[app:FreeProbability]{B} describes how the heavy-shell limit can be understood in terms of free probability. In this limit, different shell operators are relatively free (in a technical sense) with respect to each other. Appendix \hyperref[app:FreeProbability]{C} uses this connection to provide a different derivation of the entropy of certain supersymmetric black holes.

\section{Strategy: counting states} \label{sec:countingstates}

Consider a Hilbert space $\mathcal{H}$ of dimension $D$ and a family of states $\mathsf{F}_\Omega = \lbrace \ket{\Psi_{i}}\in \mathcal{H}: i=1, \dots, \Omega\rbrace$. We will be interested in computing the dimension of the Hilbert (sub)space that these states span, $d_\Omega = \text{dim}(\text{span}\lbrace\mathsf{F}_\Omega\rbrace )$, as a function of $\Omega$. This dimension is bounded above by $D$ and by the number of states of the family,
\be\label{eq:dimfamily}
d_\Omega \leq \text{min}\lbrace \Omega,D\rbrace \,.
\ee
A direct way to compute $d_\Omega$ is to consider the $\Omega \times \Omega$ Gram matrix $G$ of overlaps between states,
\begin{equation}\label{eq:Gram}
G_{ij} = \bra{\Psi_{i}} \ket{\Psi_{j}} \qquad i,j = 1,\dots ,\Omega\,.
\end{equation}
The matrix $G$ is Hermitian and positive semidefinite by construction. Its rank encodes the dimension \eqref{eq:dimfamily} through the identity
\be 
d_\Omega = \text{rank}(G) = \Omega - \text{Ker}(G)\,.
\ee 
There are different ways to evaluate $\text{rank}(G)$. One is to identify (via, e.g., the Gram–Schmidt procedure) the number of independent {\it null states} in the family, i.e. $v_i \in \text{Ker}(G)$, for which $\sum_{i=1}^\Omega v_i \ket{\Psi_i} =0$, and then simply subtract them from $\mathsf{F}_\Omega$. A related possibility is to diagonalize $G$ and to count the number of non-zero eigenvalues. These two methods require explicit input about the entries of the Gram matrix $G$. 

An alternative approach is to extract the rank from knowledge of the moments, $G^n$. This information is conveniently encoded in the resolvent matrix
\be\label{eq:resol}
R_{ij}(\lambda)\,\equiv\,\left( \frac{1}{\lambda \mathds{1} -G}\right)_{ij} \,=\,\frac{1}{\lambda}\,\delta_{ij}+\sum\limits_{n=1}^{\infty}\,\frac{1}{\lambda^{n+1}}\,(G^n)_{ij}\,.
\ee
To read the dimension we only need the trace of the resolvent
\be\label{eq:traceresol}
R(\lambda) = \frac{\Omega}{\lambda} + \sum\limits_{n=1}^{\infty}\,\frac{1}{\lambda^{n+1}}\,\text{Tr}(G^n)\,.
\ee
A standard result in linear algebra is that the eigenvalue density of $G$ is governed by the discontinuity of the trace of the resolvent along the imaginary axis, 
\be\label{eq:densitydisc}
D(\lambda)=\lim\limits_{\epsilon \rightarrow 0}\frac{1}{2\pi i}\left(\,R(\lambda-i\epsilon)-R(\lambda+i\epsilon)\,\right)\,.
\ee
The dimension is the number of non-zero eigenvalues in this density of states
\be\label{eq:dimspan}
d_\Omega = \lim\limits_{\varepsilon\rightarrow 0^+}\int_{\varepsilon}^\infty\text{d}\lambda\, D(\lambda) \,.
\ee 
This neatly shows that to get $d_\Omega$ we only need the traces of the powers of the Gram matrix. Therefore, this method is well suited to a statistical approach to compute $d_\Omega$ when the information about the matrix entries of individual instances of $G$ is hard to obtain (and possibly uninteresting). Namely, instead of the exact moments $\text{Tr}(G^n)$, we can use the averaged moments $\overline{\text{Tr}(G^n)}$ of a statistical distribution of matrices. For black hole microstates, the GPI will make this statistical approach a necessity.

Let us then assume that the family of pure states $\mathsf{F}_\Omega$ is constructed by some general protocol. For example, one can take instances of some probability distribution on the Hilbert space $\mathcal{H}$. In practice, this probability distribution can be attributed to an intrinsic source of error when preparing the microscopic states. {\it Any} such smooth probability distribution on $\mathcal{H}$, no matter how concentrated the measure is, will yield the saturation of \eqref{eq:dimfamily} with probability one, namely
\be\label{eq:saturation}
d_\Omega = \text{min}\lbrace \Omega,D\rbrace \,.
\ee
The reason is simple: any proper linear subspace generated by $\Omega < D$ states is a measure-zero subset of $\mathcal{H}$. Therefore the probability of having linear dependence is essentially zero, until the number of states is given by the Hilbert space dimension $D$. We remark again that this transition does not depend on the probability distribution used to generate the states. 

These general observations suggest a concrete strategy to obtain the Hilbert space dimension $D$ by counting states, valid in any quantum system with finite $D$. The strategy is simple: 
\begin{enumerate}
    \item Construct a family $\mathsf{F}_\Omega$ where the number of states $\Omega$ can be as large as desired.\\[-.3cm]
     \item Compute the dimension $d_\Omega$ from the trace of the resolvent of the Gram matrix $G$ using eqs.~\eqref{eq:traceresol}, \eqref{eq:densitydisc} and \eqref{eq:dimspan}.\\[-.3cm]
    \item Obtain $D$ from the saturation value of $d_\Omega$ for large enough $\Omega$.
\end{enumerate}
In the gravitational applications below, the statistical nature of the calculation will be a consequence of introducing wormhole geometries for the moments of $G$  in step 2. Otherwise, the method is generally valid.

\section{Black hole microstates}\label{sec:BHmicrostates}

We consider quantum gravitational systems in AdS$_{d+1}$ space, with boundary $\mathbb{R}\times {S}^{d-1}$, which include black holes with arbitrarily high energies as well as different charges. Microscopically, all of these black holes are part of the Hilbert space $\mathcal{H}$ of a holographic CFT$_d$ on the spatial $S^{d-1}$. The space $\mathcal{H}$ is infinite-dimensional, so to count the entropy of a black hole we need to specify the families of states $\mathsf{F}_\Omega$ associated to that particular black hole. This motivates the following definition of black hole microstate.

A black hole in equilibrium is a physical system with fixed energy $E$ and other charges $Q_I$ ($I=1,\dots ,N_Q$) in the thermodynamic limit. As in quantum statistical mechanics, a {\it microstate} $\ket{\Psi}\in \mathcal{H}$ of this black hole is any microscopic pure state which is effectively indistinguishable from the equilibrium density matrix $\rho_{\text{eq}}$ in the thermodynamic limit,\footnote{In this limit we need not specify the ensemble for  $\rho_{\text{eq}}$ since they are all equivalent.} both at the level of the conserved charges,
\be \label{eq:weak}
\langle\Psi\vert Q_I\vert\Psi\rangle\;\rightarrow\;\textrm{Tr}(\rho_{\text{eq}}\,Q_{I})\,,
\ee
and, in a stronger way, when probed with simple operators
\be\label{eq:strong}
\begin{split}
\langle\Psi\vert \mathcal{O}(t)\vert\Psi\rangle\; \rightarrow \; \textrm{Tr}(\rho_{\text{eq}}\,\mathcal{O}(0)) \,,\hspace{.8cm}\\
\langle\Psi\vert \mathcal{O}(t)\,\mathcal{O}(0)\vert\Psi\rangle \;\rightarrow \;\textrm{Tr}(\rho_{\text{eq}}\,\mathcal{O}(t)\,\mathcal{O}(0))\,.
\end{split}
\ee

These conditions are physically sensible but rarely useful in a strongly-coupled quantum many-body system. The reason is that, in general, there is no way to obtain the right-hand sides of \eqref{eq:weak} or \eqref{eq:strong}, let alone construct the wavefunctions of microstates that satisfy these relations. Notable exceptions are large-$N$ systems with a semiclassical description in terms of effective master fields, and holographic CFTs are in this class: the master field is Einstein (super)gravity in AdS. Knowing this, the by-now very well-developed AdS/CFT technology allows to obtain the right-hand sides of~(\ref{eq:weak}) and~(\ref{eq:strong}) from semiclassical black hole geometries. 

The more recent surprise is that this same master field can also be used to manufacture sets of microstates $\ket{\Psi}$ that satisfy these relations.

\subsection{Semiclassical black hole microstates with horizons}

In AdS/CFT, a microstate $\ket{\Psi}$ of the CFT lives on a spacelike slice of the conformal boundary of AdS. All these microstates have a counterpart in the bulk, but many do not  admit a good semiclassical gravitational description. However, it has long been known that at least some black hole microstates do---and as we will see, there are indeed enormous classes of them. Semiclassically, they all look like the corresponding equilibrium black hole outside of the (apparent) horizon, with all of their additional features hidden in the black hole interior. This implies almost immediately that they satisfy \eqref{eq:weak} and \eqref{eq:strong}.

A well-known example of a microstate is the time-evolved thermofield-double state \cite{Maldacena:2001kr}. It is an explicit pure state of the boundary theory that is described in the bulk by the eternal Schwarzschild-AdS geometry. It gives clear proof that it is very well possible and consistent that a black hole microstate has a horizon (and an inner singularity) in its effective semiclassical description. The black hole interior contains a long Einstein-Rosen bridge, with a throat whose area computes the fine-grained entanglement entropy of the state. 

Other natural microstates are one-sided black holes created from the collapse of matter in a pure state. In this case, the matter that formed the black hole is hidden behind the horizon. The apparent horizon is a coarse-grained notion owing to the fact that the microstate looks like an equilibrium state when probed with simple operators in the thermodynamic limit, as \eqref{eq:strong} and \eqref{eq:weak} require. More complex probes are needed to access the semiclassical black hole interior, and this information is accessible in the microstate $\ket{\Psi}$.

These examples realize the idea that microstates with a semiclassical geometry must look like the corresponding black hole in the exterior region. They further motivate a strategy to construct larger classes of microstates: we just introduce structure in the black hole interior, while leaving the exterior, including the horizon, untouched.

\subsection{Partially Entangled Grand-canonical States (PEGS)}\label{PEGs}

We will implement the previous ideas using {\it partially entangled grand-canonical states} (PEGS), which generalize the canonical PETS (partially entangled thermal states) of \cite{Goel:2018ubv}. These states are microstates of two identical copies of a holographic CFT on a spatial sphere, $\mathcal{H} = \mathcal{H}_{L} \otimes \mathcal{H}_{R}$. They have the general form
\begin{align}\label{states}
    \ket{\Psi} = \sum_{a,b}\Psi_{ab} \ket{E_a}_{L}\otimes \ket{E_b}_{R}    
\end{align}
with wavefunction\footnote{Here we restrict to $\mathbb{Z}_2$-symmetric PEGS where the left and right features of the state are taken to be the same.}
\be\label{eq:generalPEGS} 
    \Psi_{ab} =  Z_1^{-1/2}\left(e^{-\frac{\tilde{\beta}}{2}( H - \mu_I {\hat Q}_I) } \mathcal{O}e^{-\frac{\tilde{\beta}}{2}(H - \mu_I {\hat Q}_I)}\right)_{ab}\,,
    \ee 
where $H = H_{L} = H_{R}$ is the Hamiltonian, ${\hat Q}_I$ represent the charges and $\tilde{\beta},\mu_I$ represent the inverse preparation temperature and the chemical potentials, respectively. The constant $Z_1$ is a normalization factor to impose $\text{Tr}(\Psi \Psi^\dagger) =1$ and is given by a grand-canonical two-point function
\be\label{eq:norm} 
Z_1 = \text{Tr}\left(e^{-\tilde{\beta} (H - \mu_I {\hat Q}_I) } \mathcal{O}e^{-\tilde{\beta} (H - \mu_I {\hat Q}_I)}\mathcal{O}^\dagger\right)\,.
\ee
The unnormalized PEGS  $Z_1^{1/2}\Psi_{ab}$ is prepared by cutting halfway open the Euclidean CFT path integral \eqref{eq:norm}. In these constructions, the precise microstate is then specified by the particular operator $\mathcal{O}$ and the preparation temperatures/chemical potentials $\tilde{\beta},\mu_I$.

We now select the operator $\mathcal{O}$ from the class of {\it thin-shell operators} of the holographic CFT. That is, given a conformal primary $\phi_\Delta$, with $1 \ll \Delta \ll N^2$, we consider an arrangement of $O(N^2)$ local insertions of this primary on the spatial sphere, distributed in an approximately uniform way. The resulting operator has the form 
\begin{align}\label{shello}
\mathcal{O}= \prod_\alpha \phi_\Delta^{\epsilon}(\Theta_\alpha)\,,    
\end{align}
where $\Theta_\alpha \in S^{d-1}$ and each local operator insertion has been smeared over a domain $D_\epsilon \subset S^{d-1}$ of size $\epsilon^{d-1}$, in order to regularize its energy. This operator produces the microstate \eqref{eq:generalPEGS} and admits an effective semiclassical description: it creates a spherical thin interface of dust particles close to the asymptotic boundary of AdS, at a radius $r_\infty \sim \ell^2/\epsilon$ (cf.~\cite{Anous:2016kss}). For the shell to carry charge or angular momentum, the CFT operators must carry the corresponding (R-)charges or spin quantum numbers.

\begin{figure}[ht]
    \centering
    \includegraphics[width = .45\textwidth]{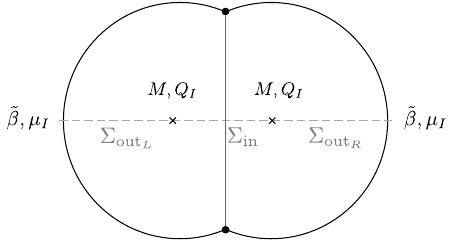}
    \caption{Euclidean saddle point bulk geometry preparing the semiclassical dual to the PEGS.}
    \label{fig:PEGSsemiprep}
\end{figure}

To prepare the semiclassical dual state in the limit $G\rightarrow 0$, we proceed \`{a} la Hartle-Hawking and cut open the Euclidean gravitational path integral that computes $Z_1$. The asymptotic Euclidean boundary conditions are set to be grand-canonical, namely, fixing the temperature, angular velocity, and holonomy of the gauge fields (see Fig.~\ref{fig:PEGSsemiprep}).

In the cases that we study in this paper, the dominant saddle point geometry contains two electrovacuum black hole solutions, which are glued in a specific way across the trajectory of the thin shell. The junction conditions impose constraints on the charges of these black holes. The preparation temperature $\tilde{\beta}$ and the chemical potentials $\mu_I$ are $ N_Q+1$ parameters. The two black holes will have the same physical properties and thus the Euclidean saddle point is also determined by $N_Q+1$ parameters, corresponding to the ADM mass $M$ of each black hole and its charges $Q_I$.

The semiclassical state is specified by initial data on a time reflection-symmetric slice $\Sigma$ of the Euclidean preparation geometry. Geometrically, the slice $\Sigma$ comprises three components: two exterior regions $\Sigma_{\text{out}_L}$ and $\Sigma_{\text{out}_R}$ delimited by two horizons, and an Einstein-Rosen bridge $\Sigma_{\text{in}}$ of non-zero length between them. The latter exists owing to the backreaction of the thin shell: without it, $\Sigma_{\text{in}}$ would shrink to the bifurcation surface of the eternal black hole. Under Lorentzian time evolution, the state evolves into a two-sided black hole with a large interior. 

These states are partially entangled since the correlations between the left and right sides decrease due to the large interior that the shell operator creates. But the exterior geometry is the same for all of them, so they are good black hole microstates according to the definition we gave above.

\subsection{PEGS with charge and rotation}\label{crm}

We now give explicit details of how to prepare the semiclassical dual to the PEGS for black holes with charge and rotation. It will become clear that the construction is very generic. To clarify how this construction can be done within String Theory, we may envisage black hole solutions in ten-dimensional type IIB supergravity with metric \cite{Chamblin:1999tk}
\begin{align}\label{ads5s5}
    ds^2=&~g_{\mu\nu}dx^\mu dx^\nu\nonumber\\
    &+\ell^2 \sum_{i=1}^3
\left(d\mu_i^2+\mu_i^2\left(d\varphi_i+\frac2{\sqrt{3}}A_\mu dx^\mu\right)^2\right)\,.
\end{align}
This ansatz describes a compactification on a sphere ${S}^5$ (with azimuthal angles $\varphi_i$ and direction cosines $\mu_i$, $\sum_i \mu_i^2=1$) down to a five-dimensional AdS spacetime with metric $g_{\mu\nu}$, $\mu,\nu=0,\dots 4$. The reduced description is a minimal five-dimensional supergravity with $SO(6)$ symmetry gauged by the Kaluza-Klein gauge potential $A_\mu$. The bosonic sector comprises the Einstein-Maxwell-AdS theory. In a general reduction, the rotation group $SO(6)$ of the $S^5$ is broken to three different $U(1)$ factors, giving a KK-reduced theory with three gauge fields $A_\mu^i$. For our purposes, though, it suffices to study the simplest situation where the three gauge fields are equal to each other.

We will consider black holes that are rotating within the AdS$_5$ factor, but notice that an electric component $A_t$ is, from the ten-dimensional viewpoint, a rotation of the $S^5$ along the three angles $\varphi_i$. The former are conventional Kerr-AdS$_5$ solutions with a rotating $S^3\subset \mathrm{AdS}_5$, while the latter give RN-AdS$_5$ black holes. In the dual CFT$_4$, i.e., the $\mathcal{N}=4$ SYM theory, the charge of these black holes is the expectation value of an R-symmetry current. 
To produce microstates with these charges in the construction of the shell operator \eqref{shello} one can use chiral primaries of the SCFT. However, the shell is then charged and one needs to combine shell-antishell operators. This can be done using the more general analysis described in appendix \ref{app:JunctionConditions}. More directly, one can also use neutral primaries, albeit those have larger scaling dimensions and do not belong to the low-energy supergravity sector \cite{Aharony:1999ti}.
A similar reduction from eleven-dimensional supergravity on $S^7/\mathbb{Z}_k$ yields an AdS$_4$ theory with a Maxwell field, whose dual is an ABJM three-dimensional gauge theory. 

The reduction to three-dimensional rotating BTZ black holes has also been well studied. Among many possibilities, one of the simplest is the compactification of type IIB String Theory on $\mathrm{AdS}_3\times S^3\times T^4$, whose dual is a SCFT with $\mathcal{N}=(4,4)$ supersymmetry. Matter lies in a hypermultiplet and a vector multiplet, and we can choose matter uncharged under the R-symmetry since the scalars in the hypermultiplet are in the trivial representation \cite{Aharony:1999ti}.

These instances serve to illustrate that the microstates that follow admit embeddings in String Theory. Nevertheless, as we will see, the construction does not require explicit knowledge of these embeddings, so we will keep it more general.

\medskip
{\bf \small Example 1: microstates with $U(1)$ charge}\medskip

Let us now build PEGS in a holographic CFT$_d$ with a $U(1)$ global symmetry that is dual to the Euclidean Einstein-Maxwell-AdS$_{d+1}$ theory
\begin{align} \label{eq:EinsteinMaxwellAction}
I_{\text{EM}} =& - \frac{1}{16 \pi G} \int_X \sqrt{g} \left( R + d(d-1) - F^2 \right) \nonumber\\
&- \frac{1}{8 \pi G} \int_{\partial X}\sqrt{h} 
K \,.
\end{align}
We use units where the AdS radius equals one. When $d=4$ this is the action that governs the IIB solutions \eqref{ads5s5}, up to a Chern-Simons term that is not relevant to the electric and static solutions that we consider.

The semiclassical description of PEGS involves a spherically symmetric thin shell. To construct it, we describe it as the trajectory
\begin{align}\label{taur}
    \tau = \mathcal{T}(T)\,,\qquad r=R(T) 
\end{align}
that is obtained by gluing two Reissner-Nordstr\"om geometries of equal mass and charge,
\be
\rd s_\pm^2 = f(r) \rd \tau^2 + f^{-1}(r) \rd r^2 + r^2 \rd \Omega_{d-1}^2 \,,
\ee
where
\be
f (r) = r^2 + 1 - \frac{16 \pi G M}{(d-1)V_{\Omega} r^{d-2}} + \frac{32 \pi^2 G^2 Q^2}{(d-1)(d-2) V_{\Omega}^2 r^{2d-4}} \,,
\ee
with $V_{\Omega} = {\rm Vol} (S^{d-1})$ and $d > 2$. The gauge potential is 
\be
A_{\pm} = i \left( \frac{4 \pi G Q}{(d-2) V_{\Omega} r^{d-2}} - \Phi \right) \rd \tau \,,
\ee
where the constant $\Phi$ is the chemical potential determined by regularity at the Euclidean horizon.

In appendix \hyperref[app:JunctionConditions]{A} we present the junction conditions along the shell, as well as a detailed example for a generic case where the black holes on each side of the shell may be different. In these general scenarios, the shell is charged and carries a non-vanishing surface current. These examples are important if one is to construct the embedding in String Theory using only fields in the low-energy supergravity sector.

Here we consider that the shell consists of neutral dust. Then the stress-energy for the shell is 
\begin{align}
    S_{ab} = - \sigma u_a u_b
\end{align}
with $u = \partial_T$, and in the $\mathbb{Z}_2$-symmetric case the junction conditions for the metric and gauge field give the equations \eqref{taur} for the trajectory of the shell. We glue two solutions with the same charges consistently with the neutrality of the shell.\footnote{In appendix \ref{app:JunctionConditions} we analyze more general scenarios with non-vanishing surface current.} The radial trajectory, $R(T)$, is governed by
 \begin{align}
     \dot{R}^2 + V_{\text{eff}}(R) = 0
 \end{align}
with
\be \label{eq:potentialCharged}
V_{\text{eff}}(R) = - f (R) + \frac{16 \pi^2 G^2 m^2}{(d-1)^2 V^2_{\Omega} R^{2d-4}} \,,
\ee
and the density $\sigma(R)$ varies to keep the mass of the shell 
\begin{align}
m = V_{\Omega} R^{d-1} \sigma(R) 
\end{align}
constant along the trajectory. 
The Euclidean time elapsed during the shell's trajectory is obtained by integrating \begin{align}
    \dot{\mathcal{T}} = f^{-1}(R) ( f(R) - \dot{R}^2 )^{1/2}\,.
\end{align}
The shell traverses the Euclidean geometry starting at the asymptotic boundary $r_{\infty}$, reaching a turning point in the bulk at the largest real root of $V_{\text{eff}}(R)$, 
\begin{align}
    r = R_{\star}\,,\qquad V_{\text{eff}}(R_{\star})=0\,,
\end{align}
and then heading back towards the boundary. In Fig.~\ref{fig:PEGSsemiprep}, the shell turns at the moment of time-reflection symmetry. The path integral over half the geometry prepares the dual to the PEGS on this Cauchy slice. This state may subsequently evolve in Lorentzian time. 

If we choose suitable preparation temperatures, the shell will always lie in the interior of the two-sided black hole. The Euclidean time it takes during its journey is
\be \label{eq:timeCharged}
\Delta \tau = 2 \int_{R_{\star}}^{r_{\infty}} \frac{\rd R}{f(R)} \sqrt{\frac{f(R) + V_{\text{eff}}(R)}{-V_{\text{eff}}(R)}} \,. 
\ee 
If the shell were absent, the periodicity of each of the full Euclidean circles would equal half the physical inverse-temperature of the black holes $\beta$, giving rise to the (Grand-Canonical) Thermofield Double.  With the shell, the preparation temperature $\tilde\beta$ is larger, 
\begin{align}
    \tilde{\beta}=\beta - \Delta \tau\,.
\end{align}
As mentioned above, we can think of each PEGS microstate \eqref{eq:generalPEGS} as characterized by three parameters: the preparation temperature and chemical potentials $\tilde\beta,\mu$, and additionally the rest mass $m$ of the CFT operator $\mathcal{O}$. However, since we want microstates dual to a given black hole with prescribed macroscopic parameters, it is more appropriate to think of $\beta$ as determined by the parameters of the black hole and then, for each shell mass $m$, to solve for $\Delta \tau$ and obtain the appropriate preparation temperature $\tilde{\beta}(m)$. The chemical potential $\mu$ is finally identified with $\Phi$. 

Two related and remarkable aspects of this construction are the following. First notice we can build geometries with shell rest mass $m$ arbitrarily larger than the ADM mass $M$ of the asymptotic regions.\footnote{For technical details see appendix \ref{app:JunctionConditions}.} The shell is trapped between two horizons and its mass is reflected in the size of the interior that it creates \cite{Balasubramanian:2022gmo}, but not in the exterior geometry. Ref.~\cite{Folkestad:2023cze} analyzes in detail how the rest mass of the shell can get screened behind a minimal codimension-two surface. These types of screenings cannot be done for the electric charge in the absence of opposite charge. Second, notice that we do not require the existence of additional degrees of freedom providing large numbers of microstates for a shell of a given mass $m$. We can create an infinite number of microstates in a very natural way, simply by varying the shell's proper mass. In the microscopic description of the PEGS, this is accomplished by varying the number of primary operator insertions of the corresponding dust-shell operator.

\medskip
{\bf \small Example 2: Rotating microstates}\medskip

A similar procedure allows to incorporate rotation. For the sake of simplicity, we start in $d=2$, where the $U(1)$ isometry is respected by the BTZ bulk solution. Working in the co-rotating frame of the shell, we get a rotating black hole on each side,
\be
\begin{aligned}
 \rd s^2 = &~f(r) \rd \tau^2 + f^{-1}(r) \rd r^2 \\
 & + r^2 \left[\rd \phi + i ( \Omega(r) - \Omega(R(\tau)) \rd \tau\right]^2 \,,
\end{aligned}
\ee
where
\be
f (r) = r^2 - 8 G M + \frac{16 G^2 J^2}{r^2} \,, \quad \Omega(r) = \frac{4 G J}{r} \,.
\ee
The junction conditions with a pressureless dust-shell require $m =2 \pi R \sigma$ to be constant along the trajectory and 
\begin{align}
    \dot{R}^2 + V_{\text{eff}}(R) = 0\,, 
\end{align}
with
\be
\begin{aligned} \label{eq:potentialBTZ}
V_{\rm eff}(R) & = - f(R) +4 G^2 m^2 \\
& = -\frac{(R^2 - R_{\star}^2)(R^2 - \tilde{R}_{\star}^2)}{R^2} \,.
\end{aligned}
\ee
In the last expression we employ the two roots of the potential, $R_{\star}$ and $\tilde{R}_{\star}$, whose form is not particularly illuminating, but they are convenient to present the solution for the shell's trajectory in the simple form
\begin{align}
    R^2(T) = \tilde{R}_{\star}^2 + (R_{\star}^2 - \tilde{R}_{\star}^2) \cosh^2 (T)\,.
\end{align}

Rotation in higher dimensions is more challenging because we lose spherical symmetry. However, rotating black holes with all angular momenta equal in odd dimension, 
\begin{align}
    D = 2 N + 3\,,
\end{align}
have an enhanced symmetry which makes the solutions non-trivially dependent on the radial coordinate only (co-homogeneity one) and thus much more tractable \cite{Myers:1986un,Emparan:2008eg, Hawking:1998kw,Gibbons:2004uw}. 

Restricting to $\mathbb{Z}_2$-invariant situations, the metric on each side of the shell has the form
\be
\begin{aligned}
\rd s^2 = & f(r)^2 \rd \tau^2 + g(r)^2 \rd r^2 + r^2 \hat{g}_{ij} \rd x^i \rd x^j \\
& + h(r)^2 \left[ \rd \psi' + A_i \rd x^i + i \Omega(r) \rd \tau \right]^2 \,,
\end{aligned}
\ee
where the $x^i$ are coordinates on $\mathbf{CP}^{N}$,  $\hat{g}_{ij}$ is the corresponding Fubini-Study metric, $A_i \rd x^i$ is the K\"ahler potential, and
\be
\begin{aligned}
g(r)^2 & = \left( 1 + r^2 - \frac{2 M (1 - a^2)}{r^{2N}} + \frac{2 M a^2}{r^{2N+2}} \right)^{-1} \,, \\
h(r)^2 & = r^2 \left( 1 + \frac{2 M a^2}{r^{2N + 2}} \right) \,, \\
f(r) & = \frac{r}{g(r) h(r)} \,, \quad \Omega(r) = \frac{2 M a}{r^{2N} h(r)^2} \,.
\end{aligned}
\ee
The mass and angular momentum of the black hole are proportional to $M$ and $M a$ respectively \cite{Kunduri:2006qa}. To work in a frame corotating with the shell we define $\psi$ such that
\begin{align}
    \rd \psi' = \rd \psi - i \Omega (R(\tau)) \rd \tau\,.
\end{align}
Since the geometry has cohomogeneity-one we can take shells that follow curves $\tau = \mathcal{T}(T)$ and $r = R(T)$, with the angular directions remaining constant.

The first junction condition requires $M a^2$ to be the same on both sides of the shell. This is automatic in the $\mathbb{Z}_2$-symmetric situation. The second condition cannot be solved if the shell is made of pressureless dust, which cannot support the rotation of the geometry. The simplest possibility in the symmetric situation (with $Ma$ equal on both sides of the shell), is to endow the shell with an anisotropic pressure,
\begin{align}
    S_{ab} = - \sigma u_a u_b + R^2 \Delta P \hat{g}_{a b}\,,
\end{align}
where $\hat{g}_{a b} = e_a^i e_b^j \hat{g}_{ij}$ is the induced metric on the shell in the directions transverse to the rotation.\footnote{More general shells were explored in \cite{Delsate:2014iia}.} Then, the dynamics of the shell is governed by the effective potential
\be \label{eq:potentialMyersPerry}
V_{\text{eff}}(R) = - g(R)^{-2} + \frac{16 \pi^2 G^2 m^2 h(R)^2}{(2N + 1)^2 V_{\Omega}^2 R^{4N + 2}} \, ,
\ee
and the density and pressure are
\be
\begin{split} \label{eq:rotationPressure}
\sigma = \frac{m}{(2N + 1) V_{\Omega} R^{4N+1}} \frac{\rd}{\rd R}\left( R^{2N} h \right) \,, \\
\Delta P = - \frac{m}{(2N+1) V_{\Omega} R^{2N}} \frac{\rd}{\rd R}\left( \frac{h}{R} \right) \,.
\end{split}
\ee

The need for $\Delta P\neq 0$ implies that the boundary insertions that create the shell cannot simply be a set of primaries characterized only by their dimension $1\ll\Delta\ll N^2$. They must carry other quantum numbers in the same range such that they produce a fluid interface that supports anisotropic pressure. It would be interesting to have explicit candidates for these insertions. Fortunately, we will see that this apparent complication is not consequential since the pressure vanishes in the universal limit of interest below.

\medskip
{\bf \small Example 3: Near-extremal black hole microstates: within or without 
the throat} 

\medskip

This construction of PEGS in a system with a $U(1)$ symmetry can be used for black holes close to extremality, where $\beta$ is large. In this regime, the two Euclidean horizons in Fig.~\ref{fig:PEGSsemiprep} develop very long throats whose geometry is very closely approximated by ${\rm AdS}_2\times S^{d-1}$. The shell created by the microstate operator $\mathcal{O}$ may or may not probe this region, and, as we will see, this can make an important distinction on the properties of the microstates.

For a given charge $Q$ of a Reissner-Nordstr\"om black hole, and in the limit where the black hole is large compared to the AdS scale, $r_0 \gg 1$, the values of the horizon radius and mass at extremality are
\be
r_0^{2d-2} \approx \frac{32 \pi^2 G^2 Q^2}{d(d-1) V_{\Omega}^2} \,, \quad r_0^d \approx \frac{8 \pi G (d-2) M_0}{(d-1)^2 V_{\Omega}} \,.
\ee
A near-extremal black hole with the same charge will be characterized by a horizon radius $r_h = r_0 + \delta r_h$ with an inverse temperature
\be 
\beta \approx \frac{2 \pi L_2^2}{\delta r_h} \,, \quad L_2^2 \equiv \frac{1}{d(d-1)} \;,
\ee 
and with mass 
\begin{align}
   M = M_0\left(1 + \frac{d(d-2)}{2} \frac{\delta r_h^2}{r_0^2}\right) \,.
\end{align}
$L_2$ is the length scale of the AdS$_2$ throat,  which appears in the metric of the near-horizon region $\rho \equiv r - r_0 \ll r_0$ (NHR) through, 
\be \label{eq:NHRMetric}
f(r) \approx \frac{\rho^2 - \delta r_h^2}{L_2^2} 
+ \mathcal{O}(\rho/r_0) \,.
\ee
This produces an ${\rm AdS}_2$ throat times a sphere of slowly varying size $r_0 + \rho$. For large charged black holes, $L_2$ is of the order of the AdS$_{d+1}$ radius and thus much smaller than the sphere radius $r_0$. This limit is convenient for simplifying calculations, but it is not necessary for our construction.

The geometry can be divided in two regions: the throat (NHR) is the region where $r-r_0\ll r_0$, and the far zone is the region where the extremal geometry is approximately valid, characterized by $r-r_0\gg \delta r_h$. The two regions overlap around the mouth of the throat, where $r-r_0=\rho_{\partial \text{NHR}}$ is such that $\delta r_h\ll \rho_{\partial \text{NHR}}\ll r_0$. 

\begin{figure}[ht]
    \centering
    \includegraphics[width = .37\textwidth]{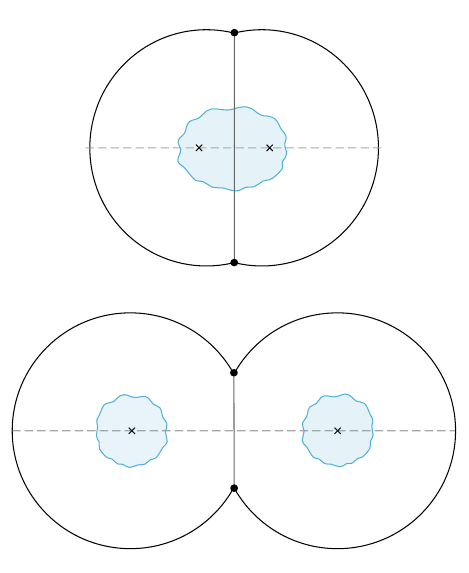}
    \caption{Different classes of shell microstates for near-extremal black holes. Above: light shells enter the near-horizon AdS$_2$ throat (light blue region). These microstates can be directly related to microstates in the JT theory that describes the throat and correspond to operator insertions in the Schwarzian theory (wiggly blue line) at the mouth of the throat. Below: In the heavy-shell limit, the shell does not probe the near-horizon region. The two-sided interior contains two AdS$_2$ throats and a weakly gravitating region where the shell resides. }
    \label{fig:twosidedschw}
\end{figure}

Whether the shell enters or not the throat depends on the value of the largest root $R_{\star}$ of the effective potential \eqref{eq:potentialCharged}: if $R_{\star}$ is larger/smaller than $r_0 + \rho_{\partial \text{NHR}}$ then the shell will remain without/within the throat. In particular, the shell avoids the throat if
\be \label{eq:shellInsideThroat}
\begin{split}
m L_2 \gg \frac{r_0^{d-1}}{G} \, . \\
\end{split}
\ee
This is depicted in Fig \ref{fig:twosidedschw}. Although eq.~\eqref{eq:shellInsideThroat} is derived assuming $r_0\gg L_2$, for small black holes with $r_0\sim L_2$ the shell also avoids the throat at sufficiently large masses. Recall also that we are always considering that the masses are heavy enough, $\mathcal{O}(1/G)\sim\mathcal{O}(N^2)$, to produce appreciable backreaction.

We can now relate our microstates to the PETS introduced  in \cite{Goel:2018ubv} in JT gravity. This theory describes the near-horizon throat, and its dynamics is localized at the mouth in the form of the Schwarzian theory of fluctuations of the boundary curve \cite{Maldacena:2016upp}. The states correspond to insertions of local operators on the boundary curve. This curve imposes an upper cutoff on the conformal dimension for the operators, above which the JT gravity description ceases to be valid. Operators of higher dimension would be beyond the cutoff, and in our construction they correspond to shell masses exceeding the lower bound in \eqref{eq:shellInsideThroat}, for which the shell's trajectory escapes the ${\rm AdS}_2$ region. 

By working in a higher-dimensional setup where we keep the region of spacetime outside the throat we can increase the mass of the shell in the range \eqref{eq:shellInsideThroat}. Indeed we can make it arbitrarily large.\footnote{The mass of the shell is then related to the dimension of an operator of the $d$-dimensional CFT, instead of the property of an operator of the quantum mechanical system describing the throat.} Although one loses the large degree of calculability afforded by the simplicity of the dynamics in the throat, we will see that by taking a limit of very heavy shell masses we achieve a crucial feature of the construction: its universality.

Finally, we remark again that this construction also holds in supersymmetric scenarios, e.g. those arising in type IIB in ${\rm AdS}_5$ or the four-dimensional ungauged $\mathcal{N}=2$ supergravity in asymptotically flat space. In particular, it provides microstates for its extremal BPS solutions. The reason is that the bosonic part of these theories is still Einstein-Maxwell theory. We will return to this point in Sec.~\ref{sec:corrections}.

\subsection{Microstates of asymptotically flat black holes}\label{subsec:AF}

It should be apparent that the shell construction can be readily adapted to black holes in asymptotically flat geometries (or indeed almost any other asymptotics) \cite{Balasubramanian:2022lnw}. We need not redo the calculations, but simply consider the limit where the black hole is much smaller than the AdS radius. All that follows, in particular the derivation of the dimension of the Hilbert space of shell microstates and the universality of the result, carry through at a technical level. This is a significant advantage of the shell construction of microstates.

However, there are two important conceptual caveats to bear in mind. The first one is that, lacking sufficient understanding about asymptotically flat holography, it is unclear what is the dual microscopic counterpart of the shell in the bulk. It seems very plausible that, if a holographic dual exists, then it should allow for the construction of bulk shells, but we cannot be any explicit about it.

The second issue is that the canonical and grand-canonical ensembles are not properly defined in asymptotically flat space. The black holes are thermodynamically unstable systems and are not proper saddle points of the GPI, since the fluctuations around them include a Euclidean
negative mode \cite{Gross:1982cv}. The same problem affects small AdS black holes. However, this is also a feature of the Gibbons-Hawking approach, which nevertheless manages to correctly capture a surprising amount of the thermodynamics of black holes with almost any kind of asymptotics. 
Our construction shares all the advantages and shortcomings of the Gibbons-Hawking use of the GPI for the study of black holes. We adopt the position that, although asymptotically flat black holes are not described by stationary (grand-canonical) density matrices, the question of how many linearly independent black hole microstates exist is still sensible and well-defined. This question can be regarded as posed in a microcanonical ensemble, and it is the one that our approach provides an answer to.

\section{Counting microstates}\label{sec:countingmicro}

The previous constructions of black hole microstates easily yield large families of them. The physical property that we rely on is that the ADM mass $M$ and charges $Q_I$ of the exterior black holes do not depend on the rest mass $m$ of the thin shell in the interior. Therefore, we can keep $M$ and $Q_I$ fixed, or $\beta$ and $\mu_I$ if we impose grand-canonical conditions, and define multiple microstates of the same black hole, $\ket{\Psi_{m}}$, by simply varying the rest mass $m$. These semiclassical states share the same exterior geometry but differ in their interiors. Selecting a discrete and finite set of values for the rest mass then defines the family
\be\label{eq:familyPEGS}
\mathsf{F}_\Omega = \lbrace \ket{\Psi_{m_1}},\dots, \ket{\Psi_{m_\Omega}}\rbrace \subset \mathcal{H}\,.
\ee

\subsection{Microcanonical family}

Families of grand-canonical PEGS wavefunctions $\Psi_{ab}$ come with a difficulty. They have tails at all energies and therefore explore the full infinite-dimensional Hilbert space of the CFT. Assuming the genericity of the wavefunctions \eqref{eq:generalPEGS}, these states will all be linearly independent and we will have $d_\Omega = \Omega$ for any finite family of PEGS. 

This problem and its resolution are well-known in quantum statistical mechanics. The entropy $S$ of the black hole can only be understood in a microcanonical sense, as the number of independent microstates of the black hole in a window of energy 
\begin{align}
    E \in [E_\alpha,E_\alpha +\Delta E_\alpha]\,,\qquad \Delta E_\alpha \ll E_\alpha\,,
\end{align}
and charge
\begin{align}
    Q_I \in [Q_I^\alpha, Q^\alpha_I+\Delta Q^\alpha_I]\,,\qquad \Delta Q^\alpha_I \ll Q^\alpha_I\;. 
\end{align}
To lighten the presentation we omit the angular momentum, but it should be clear that all we say about charge extends to spin. The dimension $D_\alpha$ of this microcanonical Hilbert subspace $\mathcal{H}_\alpha \subset \mathcal{H}$ is finite, and gives the entropy of the black hole as 
\begin{align}
    S=\log D_\alpha\,.
\end{align}

In order to obtain $S$ by counting states of the family of PEGS we will use their projections
\be 
\ket{\Psi_{m_i}^{\alpha}} = \left(\dfrac{Z_{1,m_i}}{\mathbf{Z}_{1,m_i}^{\alpha}}\right)^{1/2}\,\Pi_{\alpha} \ket{\Psi_{m_i}}\,,
\ee
where $\Pi_{\alpha} = \Pi_{\alpha}^L\Pi_{\alpha}^R$ is an orthogonal projector onto the microcanonical band labeled by the parameter\footnote{For two-sided states, the microcanonical energy and charges need to be specified independently on both sides. This is implicitly defined in our notation, where explicitly $E_\alpha = (E^L_\alpha,E^R_\alpha)$ and $Q^\alpha_I = (Q^\alpha_{I,L},Q^\alpha_{I,R})$. The orthogonal projector factorizes accordingly, $\Pi_\alpha = \Pi_{\alpha}^L\Pi_{\alpha}^R$. We have omitted this for ease of exposition.} 
\begin{align}
    \alpha = (E_\alpha,Q^\alpha_I) \,,
\end{align}
and $\Pi_{\alpha}^L,\Pi_{\alpha}^R$ correspond to the respective projections acting on each Hilbert space factor. The function
\begin{align}\label{eq:normalizationmicro}
    \mathbf{Z}_{1,m_i}^{\alpha} = Z_{1,m_i}\bra{\Psi_{m_i}}\Pi_{\alpha} \ket{\Psi_{m_i}}\,,
\end{align}
is the normalization of the microcanonical state. Using \eqref{eq:generalPEGS}, the state vector is
\begin{align}
    \ket{\Psi_{m_i}^\alpha} = \sum_{a,b}(\Psi_{m_i}^\alpha)_{ab} \ket{E_a}_{L}\otimes \ket{E_b}_{R}\,,
\end{align}
where the sum is implicitly restricted to energies and charges that belong to the window $\alpha$. Explicitly, the wavefunction reads\footnote{Here we are using the implicit notation $\tilde{\beta}( E_\alpha - \mu_I Q_I^\alpha) \equiv \tilde{\beta}( E^L_\alpha + E^R_\alpha  - \mu_I Q_{I,L}^\alpha- \mu_I Q_{I,R}^\alpha)$.}
\be\label{eq:microcanonicalPEGS} 
(\Psi^\alpha_{m_i})_{ab} \approx  \frac1{\sqrt{\mathbf{Z}_{1,m_i}^{\alpha}}} e^{-\frac{\tilde{\beta}}{2}( E_\alpha - \mu_I Q_I^\alpha)} (\mathcal{O}^i_\alpha)_{ab}\,,
\ee 
and the normalization \eqref{eq:normalizationmicro} is
\begin{align}
    \mathbf{Z}_{1,m_i}^{\alpha} \approx e^{-\tilde{\beta}( E_\alpha - \mu_I Q_I^\alpha)} \text{Tr} ( \mathcal{O}^i_\alpha \mathcal{O}^i_\alpha {}^\dagger)\,,
\end{align}
 where $\mathcal{O}^i_\alpha = \Pi^L_\alpha \mathcal{O}^i\Pi^R_\alpha$ is the projection of the operator, and $\mathcal{O}^i$ is the microscopic thin-shell operator of rest mass $m_i$. After the projection, in these expressions we have approximated the smooth part of \eqref{eq:generalPEGS} by the corresponding constant values on the microcanonical window. This leads to the wavefunction
 \be 
(\Psi^\alpha_{m_i})_{ab} \approx  \dfrac{(\mathcal{O}^i_\alpha)_{ab}}{\text{Tr} (\mathcal{O}^i_\alpha \mathcal{O}^i_\alpha {}^\dagger)^{1/2}} \,.
 \ee 
Then, from a grand-canonical family of PEGS \eqref{eq:familyPEGS}, we can define the associated microcanonical family
\be 
\mathsf{F}^\alpha_\Omega = \lbrace \ket{\Psi^{\alpha}_{m_1}}, \dots, \ket{\Psi^{\alpha}_{m_\Omega}}\rbrace \subset \mathcal{H}_\alpha\,.
\ee 

To find the dimension $D_\alpha$ of $\mathcal{H}_\alpha$ we will follow the strategy outlined in Sec.~\ref{sec:countingstates}. It can be implemented in two manners, which differ in the way they obtain the traces of the moments of the microcanonical Gram matrix 
\be\label{eq:PEGSmicrogm}
G^{\alpha}_{ij} \equiv \bra{\Psi^{\alpha}_{m_i}}\ket{\Psi^{\alpha}_{m_j}} \approx \dfrac{\text{Tr} (\mathcal{O}^j_\alpha \mathcal{O}^i_\alpha {}^\dagger)}{(\text{Tr} (\mathcal{O}^i_\alpha \mathcal{O}^i_\alpha {}^\dagger)\text{Tr} (\mathcal{O}^i_\alpha \mathcal{O}^i_\alpha {}^\dagger))^{1/2}}\,.
\ee 
In the first method, these are extracted from those of the PEGS Gram matrix
\be\label{eq:PEGSgm}
G_{ij} \equiv \bra{\Psi_{m_i}}\ket{\Psi_{m_j}}\,.
\ee  
This was the approach used in \cite{Balasubramanian:2022gmo}, and we will work it out in detail for the cases with charge and rotation that we introduced above. The second method directly computes $G^{\alpha}_{ij}$ and its moments generalizing a prescription recently put forward in \cite{Chua:2023srl}. To avoid excessive repetition, in this last method we will only outline the calculations.

\subsection{Method 1: projected overlaps}

Inserting the resolution of the identity, $\mathsf{1} = \sum_\alpha \Pi_{\alpha}$, the matrix $G$ is written as the weighted sum over matrices $G^\alpha$, namely
\be\label{eq:microcanPEGSgm}
G_{ij} = \dfrac{1}{\sqrt{Z_{1,m_i}Z_{1,m_j}}}\sum_\alpha G^{\alpha}_{ij} \sqrt{\mathbf{Z}_{1,m_i}^\alpha \mathbf{Z}_{1,m_j}^\alpha}\,.
\ee
Recall that, from \eqref{eq:generalPEGS}, the Gram matrix of PEGS \eqref{eq:PEGSgm} is
\be 
G_{ij} = \frac{\text{Tr}\left(e^{-\tilde{\beta}_{ij} (H - \mu_I Q_I) } \mathcal{O}^je^{-\tilde{\beta}_{ij} (H - \mu_I Q_I)}\mathcal{O}^i{}^\dagger\right)}{(Z_{1,m_i}Z_{1,m_j})^{1/2}} \,,
\ee 
where $\tilde{\beta}_{ij} = \frac{\tilde{\beta}_{i} + \tilde{\beta}_{j}}{2}$. After computing $G$ and its products using the GPI, we will approximate the sums over energy windows $\alpha$ by continuous integrals. An inverse Laplace transform then gives the microcanonical quantities.

\medskip
\textbf{\small i) Universal overlaps from heavy-shell wormholes}\medskip

The trace of the resolvent matrix \eqref{eq:traceresol} contains the products
\be \label{momentG}
G_{i_1i_2}G_{i_2i_3}\dots G_{i_n i_1}\,.
\ee 
In principle, these are fully microscopic quantities, and their computation requires details about the UV theory of quantum gravity that the semiclassical path integral of gravity does not contain. However, much less information is needed if we only want to obtain the rank $d_\alpha$ of $G^\alpha$. As we discussed, a statistical averaging of the values of \eqref{momentG}---the moments of a distribution of matrices---suffices for that. The idea (exploited in  \cite{Stanford:2020wkf,Sasieta:2022ksu, deBoer:2023vsm,Balasubramanian:2022gmo,Balasubramanian:2022lnw,Penington:2019kki,Boruch:2023trc,Hsin:2020mfa,Antonini:2023hdh} in different scenarios) is that the GPI
can provide this information if appropriately used.\footnote{Perhaps the GPI, complemented with additional UV ingredients, might go beyond the statistical description and compute the microscopic inner products, at least in low-dimensional toy models \cite{Marolf:2020xie,Maxfield:2023mdj}. We will nevertheless assume that the GPI is maximally agnostic about the microscopic phases of the inner products \cite{deBoer:2023vsm}.}

We will employ an overline to denote the quantities that the semiclassical GPI gives,
\begin{align}\label{olinemomentG}
    \overline{G_{i_1i_2}G_{i_2i_3}\dots G_{i_n i_1}}\,.
\end{align}
Naively, the individual elements $\overline{G_{ij}}$ would have all the available information about the overlap between microstates. However, as we will presently explain, for a very generic family of microstates the value of $\overline{G_{ij}}$ does not contain \emph{any} information about their overlaps, making it seem that the GPI can hardly give a number $D=e^S$ that is consistent with the Gibbons-Hawking calculation of the entropy. Indeed, if we only consider $\overline{G_{ij}}$ and its products, the result for the Hilbert space dimension is infinite, which is a version of the bags-of-gold paradox \cite{Balasubramanian:2022gmo,Wheeler}.

\begin{figure}[ht]
    \hspace{.2cm}
    \includegraphics[width = .21\textwidth]{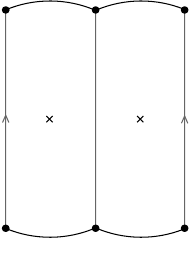}
    \hspace{.7cm}
    \includegraphics[width = .19\textwidth]{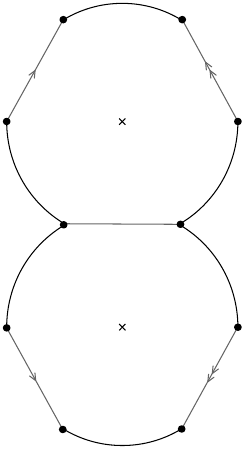}
    \hspace{.2cm}
    \caption{The two- and three-boundary wormhole contributions to the moments of the Gram matrix. Trajectories with arrows are identified. For general $n$, the solution consists of two Euclidean black holes glued along $n$ trajectories. The Euclidean solution contains $n$ disconnected asymptotic boundaries, each of which corresponds to an overlap. Each shell is taken to be different, so this solution is the only possible contraction of the operators.}
    \label{fig:wh}
\end{figure}

However, the GPI is a smarter tool than that. Connected wormhole geometries with multiple boundaries (see Fig.~\ref{fig:wh}) give contributions such that\footnote{To lighten the notation we have redefined $Z_{1,m_{i_j}}\rightarrow Z^{i_j}_1$.}
\be \label{eq:OverlapsNonUniversal}
\overline{G_{i_1i_2}G_{i_2i_3}\dots G_{i_n i_1}} = \dfrac{Z_{n}^{i_1\dots i_n}}{Z^{i_1}_1\dots Z^{i_n}_1}\,,
\ee 
for distinct $i_1,\dots,i_n$. These differ from the naive product $\overline{G_{i_1i_2}}\,\overline{G_{i_2i_3}}\dots\overline{G_{i_n i_1}}$, implying that \eqref{eq:OverlapsNonUniversal} must be regarded as the statistical moments of a distribution. This statistical nature of the GPI was first recognized in \cite{Saad:2019lba} for the correlations of partition functions in JT gravity.
It turns out to be consistent with a maximally ignorant description: semiclassics captures the smooth `phase-correlated' part of the microscopic moments, while the phases of individual matrix elements of the microscopic Gram matrix are washed out \cite{Sasieta:2022ksu, deBoer:2023vsm}. Remarkably, the smooth part of the moments is enough to obtain $d_\Omega$. 

Following \cite{Balasubramanian:2022gmo,Balasubramanian:2022lnw}, we will make the family of microstates as generic as possible by taking them to be at an infinite distance from each other in the semiclassical phase space. More concretely, we require the ``heavy-shell'' limit
\begin{align}\label{heavyshell}
m_i \rightarrow \infty\,,\quad \textrm{with}\quad |m_i-m_j| \rightarrow \infty\,,   
\end{align}
for all members of $\mathsf{F}_\Omega$.\footnote{The limit must hold besides the fact that the $m_i$ and their differences are all $O(N^2)$.} 
Then, the semiclassical interiors for different states are as different as possible, since they contain shells with infinitely many different particles relative to one another. We emphasize that this choice is just a technical convenience. As explained in the section \ref{sec:countingstates}, any set drawn from some smooth probability distribution will suffice to provide the correct Hilbert space dimension. It would be interesting to verify this for families of shell states lighter than \eqref{heavyshell}.

The heavy-shell limit brings in drastic simplifications. First, it yields
\be \label{unitG}
\overline{G_{ij}} = \delta_{ij}\,,
\ee 
so the semiclassical states would seemingly become orthogonal.\footnote{The semiclassical theory has a naive infinite Hilbert space in the black hole interior, since the slice $\Sigma_{\text{int}}$ can have an arbitrarily large volume. Therefore, the states can be made as orthogonal as one wants, which is achieved in the heavy-shell limit.} This conclusion would be too quick, since wormhole contributions give a non-zero value for the moments \eqref{eq:OverlapsNonUniversal}. In the heavy-shell limit \eqref{heavyshell} they take a very simple form, 
\be\label{eq:uniwormholen} 
\overline{G_{i_1i_2}G_{i_2i_3}\dots G_{i_n i_1}} = \dfrac{Z(n\beta;\mu_I)^{2}}{Z(\beta;\mu_I)^{2n}}\,,
\ee 
where $ \log Z(n\beta;\mu_I)$ is minus the Euclidean action of the corresponding black hole. The physical mechanism leading to these universal overlaps is that, in the heavy-shell limit, the wormhole geometry becomes a pair of Euclidean black holes effectively glued together at a small portion of the asymptotic region (which lies inside the black hole). The intrinsic contribution of the shell action, which only explores the asymptotic region, factors out and cancels between the wormhole and the normalization of the PEGS. This leads to the ratio of Gibbons-Hawking actions in \eqref{eq:uniwormholen}. 

Remarkably, the moments \eqref{eq:uniwormholen} only depend on the parameters that specify the black hole. All the information about specific microstates, such as the shell masses and insertion angles, has been rendered irrelevant. Moreover, this form of the overlaps is valid for any black hole solution for which the Euclidean action can be computed. Then, the dimension of the Hilbert space of these microstates is universally determined by the value of the Euclidean black hole action. Conceptually, this is an entirely different computation of the black hole entropy than in the Gibbons-Hawking calculus. Indeed there is no reason a priori to expect that both results match with each other. But they do, as we show in the next section.

To make manifest the mechanism behind this crucial result---and indeed to show that it implies $D_\alpha =e^{A/4G\hbar}$---we will examine in detail how it is realized in the explicit examples that we introduced above.

\medskip
{\bf \small Example 1: $U(1)$ charge}\medskip

We have to evaluate the action \eqref{eq:EinsteinMaxwellAction} for the different geometries involved in the overlaps. The first one is the normalization factor, represented in Fig.~\ref{fig:PEGSsemiprep}. To the spacetime action we must add the appropriate counterterms $I_{\text{ct}}$ and a contribution from the shell,
\be
I_{\text{shell}} = \int_{\mathcal{W}} \left( \sigma + A_{\mu} j^{\mu} \right) \,.
\ee
The full action is then $I = I_{\text{EM}} + I_{\text{shell}} + I_{\text{ct}}$, and we set $j^{\mu} = 0$ since in the $\mathbb{Z}_2$-symmetric situation we deal with neutral shells.

To compute the action, it is useful to split the geometry in Fig.~\ref{fig:PEGSsemiprep} in three regions: two circular sectors $X_{\text{L}}$ and $X_{\text{R}}$ centered at the horizons and with opening Euclidean time $\tilde{\beta}$, and the remaining interior region containing the shell $X_{\text{in}}$. The first two pieces just give contributions proportional to the grand-canonical free energy,
\be
\begin{aligned}
I[X_{\text{L}, \text{R}}] = & \tilde{\beta} \mathcal{F}(\beta, \Phi) \\
= & \frac{\tilde{\beta} V_{\Omega}}{16 \pi G} \left( -r_h^d + r_h^{d-2} \left( 1 -  \frac{2(d-2)}{d-1}\Phi^2 \right) + c_d \right) \,,
\end{aligned}
\ee
where $r_h$ is the horizon radius, and $c_d$ accounts for the Casimir energy of the CFT in even dimensions. The contribution from $X_{\text{in}}$ is more complicated but it simplifies in the heavy-shell limit. In that case, the turning point of the trajectory can be obtained as the largest root of \eqref{eq:potentialCharged},
\be \label{eq:ReissnerNordstromTurningPoint}
R_{\star}^{d-1} \approx \frac{4 \pi G m}{(d-1) V_{\Omega}} \,.
\ee
Also, the time elapsed by the shell \eqref{eq:timeCharged} scales as 
\begin{align}
\Delta \tau \sim R_{\star}^{-1} \sim m^{-1/(d-1)}\,.  
\end{align}
This confirms that in the large $m$ limit the shell stays within the asymptotic region (inside the black hole), traveling for an arbitrarily short amount of Euclidean time. As a consequence, in this limit we have 
\begin{align}
\tilde{\beta} \approx \beta\,.     
\end{align}
Using the on-shell condition
\be
R = - d(d+1) +\frac{d-3}{d-1}F^2 + \frac{16 \pi G}{d-1} \sigma \, \delta_{\mathcal{W}}(x) \,,
\ee
to compute the Einstein-Hilbert action, the contribution from $X_{\text{in}}$ becomes
\be \label{eq:onShellActionShell}
I[X_{\text{in}}] \approx \frac{2 m}{d-1} \log (2 r_{\infty}^{d-1}) - 2 m \log R_{\star} \,.
\ee
The first piece is removed with a counterterm at the shell insertions. Therefore, the normalization for the shell states in this limit is finally
\be \label{eq:shellNormalization}
Z^i_1 = Z(\beta; \Phi)^2 e^{-2 m_i \log R_{\star}^{(i)}} \,.
\ee

The overlaps between states are computed by the wormholes sketched in Fig.~\ref{fig:wh}. When $m$ is large, the length of each asymptotic segment between two boundary insertions is equal to $\beta$. The action is computed through a decomposition of the geometry like we have done for the normalization. Isolating the contributions from the shells becomes simple when these are heavy, each of them contributing $2 m_i \log R_{\star}^{(i)}$ to the action. The remaining disk sectors around each of the two horizons have an opening Euclidean time $n \beta$ for an $n$-boundary wormhole. We conclude that in the heavy-shell limit
\be \label{eq:onShellActionWormhole}
Z_n^{i_1 \dots i_n} \approx Z(n \beta; \Phi)^2 \exp \left( - 2 \sum_{i=1}^n m_i \log R_{\star}^{(i)} \right) \,.
\ee
Combined with \eqref{eq:shellNormalization}, this gives the universal result \eqref{eq:uniwormholen}.

\medskip
{\bf \small Example 2: rotation}\medskip

We can repeat this procedure for the rotating BTZ black hole. The chemical potential is now the angular velocity of the horizon $\omega = r_i / r_h$, with $r_i$ and $r_h$ the inner and outer horizon radii. The action of the disk sectors excluding the shell term gives again the grand-canonical free energy of the black hole
\be
I[X_{\text{L},\text{R}}] = \tilde{\beta} \mathcal{F}(\beta, \omega) = - \frac{\pi^2 \tilde{\beta}}{2 G \beta^2 (1 - \omega^2)} \,.
\ee
The geometry is simple enough that the contribution from $X_{\text{in}}$ can be computed analytically using the form of the shell's trajectory. But again the heavy-shell limit gives a cleaner and more illuminating result. In this limit, the roots of the potential \eqref{eq:potentialBTZ} and the Euclidean time elapsed by the shell simplify to 
\begin{align}
    R_{\star} \approx 2 G m\,,\quad \tilde{R}_{\star} \approx 2 J / m\,, \qquad \Delta \tau \approx (G m)^{-1}\,.
\end{align}
As expected, the shell localizes near the boundary, where it spends a vanishingly short Euclidean time. Its contribution to the action is again of the form
\be
I[X_{\text{in}}] \approx 2 m \log (2 r_{\infty}) - 2 m \log R_{\star} \,.
\ee 
After counterterm subtraction, we find that \eqref{eq:shellNormalization} gives the normalization of BTZ shell microstates, with the appropriate chemical potential $\Phi \to \omega$.

The computation of overlaps from wormholes follows the same lines as in the charged case, and the cancellation of contributions from the shells yields again the universal result \eqref{eq:uniwormholen}.

Essentially the same analysis is valid for rotating black holes with all angular momenta equal in $D = 2N + 3$, whose microstates were described above in \ref{crm}. The sectors $X_{\text{L},\text{R}}$ of the geometry produce a contribution proportional to the free energy. Again the heavy-shell limit allows to obtain simple and explicit results for the shell's trajectory. The turning point of the potential \eqref{eq:potentialMyersPerry} becomes
\begin{align}
    R_{\star}^{2N+1} \approx \frac{4 \pi G m}{(2N + 1) V_{\Omega}}\,, 
\end{align}
and the travel time of the shell is suppressed as
\begin{align}
    \Delta \tau \sim R_{\star}^{-1} \sim m^{-1/(2N+1)}\,.
\end{align}

The need for the anisotropic pressure $\Delta P$ in \eqref{eq:rotationPressure} generically complicates the evaluation of the action in $X_{\text{in}}$. But in the heavy-shell limit its contribution
\be
\int_{\mathcal{W}} \Delta P \approx \frac{4N + 4}{2N + 1} \frac{\sqrt{\pi} \Gamma(\frac{N+1}{2N+1})}{\Gamma(\frac{1}{4N+2})} \frac{m M \ell a^2}{R_{\star}^{2N+2}} \sim \frac{1}{R_{\star}} 
\ee
is suppressed.
The reason is that the shell passes very close to an asymptotic region hidden inside the wormhole, where the effects of rotation become negligible. 

As a result, we recover \eqref{eq:onShellActionShell} for the action of $X_{\text{in}}$, and the $n$-boundary wormhole contribution is given by \eqref{eq:onShellActionWormhole}, where the chemical potential $\Phi$ for rotation is the angular velocity. Both combined, this produces again the universal result \eqref{eq:uniwormholen}.

\medskip
\textbf{\small ii) Inverse Laplace transform}\medskip

The previous calculations have given us the universal moments of $G_{ij}$ \eqref{eq:uniwormholen}. It is now a simple matter to derive from them the moments of the microcanonical $G^\alpha_{ij}$ defined in \eqref{eq:PEGSmicrogm}, whose universal form is also remarkably simple.

In the heavy-shell limit, it is natural to assume that the individual matrices $(O^i_\alpha)_{ab}$ become effectively free relative to each other (see Appendix \ref{app:FreeProbability}). The phase-correlated moments of $G^\alpha$ are therefore of the form
\begin{equation}
\overline{G^\alpha_{i_1 i_2}G^\alpha_{i_2 i_3}\dots G^\alpha_{i_n i_1}}\,  =  \frac{\textbf{Z}^\alpha_n}{\textbf{Z}^\alpha_{1,i_1}\dots \textbf{Z}^\alpha_{1,i_n}}\,,
\end{equation}
where
\begin{align}\label{Znalpha}
    &\mathbf{Z}_{n}^{\alpha} = e^{-n\beta( E_\alpha - \mu_I Q_I^\alpha)} f^\alpha_n\,,\\
    &f^\alpha_n =\overline{\text{Tr} (\mathcal{O}^{i_1}_\alpha \mathcal{O}^{i_2}_\alpha{}^\dagger)...\text{Tr}(\mathcal{O}^{i_n}_\alpha \mathcal{O}^{i_1}_\alpha{}^\dagger)}\,.
\end{align}
The $f^\alpha_n$ functions encode the information about the overlaps and we must extract them from our previous GPI calculations. As we will now see, these functions become $i$ and $n$ independent in the heavy-shell limit, which can be motivated from the point of view of free probability for the $(\mathcal{O}^{i}_\alpha)_{ab}$ matrices (see Appendix \ref{app:FreeProbability}).

In the same way, the phases of $G^\alpha$ for different microcanonical windows $\alpha$ can be assumed to be uncorrelated, so their semiclassical connected amplitudes vanish. Then, \eqref{eq:microcanPEGSgm} implies that the moments of $G$ are simply sums of the moments of $G^\alpha$,
\begin{equation}
\overline{G_{i_1 i_2}G_{i_2 i_3}\dots G_{i_n i_1}}\,  =  \dfrac{\sum_{\alpha }\textbf{Z}^\alpha_n}{{Z}_{1,i_1}\dots {Z}_{1,i_n}}\,.
\end{equation}
The result \eqref{eq:OverlapsNonUniversal} for the lhs of this equation implies that the numerator of the rhs is the value of $n$-boundary wormhole action. In the limit of heavy-shell microstates, this action takes the universal form \eqref{eq:uniwormholen} and we find\footnote{In the heavy-shell limit, the value of $\mathbf{Z}_{n}^{\alpha}$ factorizes into $(\mathbf{Z}_{n}^{\alpha})_L(\mathbf{Z}_{n}^{\alpha})_R$, from the assumptions of randomness and independence of the $(\mathcal{O}^{i}_\alpha)_{ab}$ coefficients. This makes $\sum_\alpha \mathbf{Z}^\alpha_n$ the square of some one-sided quantity. For ease of exposition, we will implicitly restrict the sum over $\alpha$ as single-sided. This gets rid of the square in the numerator of \eqref{eq:uniwormholen}.}
\be \label{Zndiscr}
Z(n\beta;\mu_I)=\sum_\alpha \mathbf{Z}^\alpha_n\,.
\ee 
To invert this relation and extract the $\mathbf{Z}^\alpha_n$ we first use that in a semiclassical regime the sums are well approximated by continuous integrals. Inserting \eqref{Znalpha} we write \eqref{Zndiscr} as
\be 
Z(n\beta;\mu_I)= \int \text{d}E\, \text{d}Q_I\, e^{-n\beta( E - \mu_I Q_I) } f_n(E,Q_I)\,.
\ee 
An inverse Laplace transform readily yields the microcanonical $f_n(E,Q_I)$,
\begin{align}
    f_n(E,Q_I) = z(E,Q_I)\equiv \int_{\gamma +i \mathbb{R}} \dfrac{\text{d}{\beta}\,\text{d}\mu_I}{2\pi i}e^{{\beta} (E-\mu_I Q_I)} Z(\beta;\mu_I)\,,
\end{align}  
that is, $f_n(E,Q_I)$ is the microcanonical density of states $z(E,Q_I)$
that one obtains as the inverse Laplace transform of the Gibbons-Hawking partition function. This density of states does not arise from a proper counting of states, but its logarithm,
\begin{align}
    \Sbh \equiv \log z(E,Q_I)\,,
\end{align}
will presently acquire the interpretation of the entropy in the true counting of the set of microstates that we have constructed. At this stage, $S$ is merely a quantity that characterizes the moments of the microcanonical matrix of overlaps,\footnote{We omit the dependence on the width of the microcanonical window, which can be made small enough with an appropriate choice of $\Delta E_\alpha$ and $\Delta Q_I$, e.g., $\beta \Delta E_\alpha \sim O(S^0)$.}
\begin{equation}\label{microuniov}
\overline{G^\alpha_{i_1 i_2}G^\alpha_{i_2 i_3}\dots G^\alpha_{i_n i_1}}\,  =  e^{-(n-1)\Sbh}\,.
\end{equation}
With this result, it will be an easy step to deduce in Sec.~\ref{subsec:dimH} that the dimension of the Hilbert space is indeed $e^S$. But before we do this, we will examine another, more direct method to derive \eqref{microuniov}.

\subsection{Method 2: direct overlaps}

As an alternative to the previous method, the microcanonical Gram matrix and its moments can be directly computed using an extension of the semiclassical prescription put forward in \cite{Chua:2023srl} (first proposed in JT gravity in \cite{Harlow:2018tqv}). There it was shown that the microcanonical wavefunction of the Hartle-Hawking state can be obtained as the on-shell action of the `pac-man' geometry, where one imposes fixed-energy and charge boundary conditions on a bulk hypersurface $\Sigma'$ such that $\partial \Sigma'$ is the boundary slice where the microstate lives (in this case, the thermofield-double). The fixed-energy boundary conditions imposed on $\Sigma'$ are of mixed type: certain components of the induced metric and the extrinsic curvature of $\Sigma'$ are specified. Moreover, $\Sigma'$ is allowed to have corners, where Hayward terms are added to ensure the additivity of the gravitational on-shell action. This is important to ultimately recover the norm of the Hartle-Hawking state $Z(\beta;\mu_I)$ from the microcanonical wavefunction.

We will generalize this prescription to the wormhole calculus in order to compute moments of the microcanonical Gram matrix \eqref{eq:PEGSmicrogm}
\be \label{microG}
G^\alpha_{ij} = \left(\dfrac{Z_{1,i}Z_{1,j}}{\mathbf{Z}_{1,i}^{\alpha}\mathbf{Z}_{1,j}^{\alpha}}\right)^{1/2}\sum_{a} \bra{\Psi_{m_i}}\ket{E_a}\bra{E_a}\ket{\Psi_{m_j}}\,,
\ee 
where, again, the sum is implicitly restricted to energies and charges that belong to the window $\alpha$. We will focus on the overlap between the (unnormalized) thin-shell microstates and energy eigenstates,
\begin{align}
    \Psi_i(E) \equiv Z_{1,i}^{1/2}\bra{E}\ket{\Psi_{m_i}}\,.
\end{align}

It is immediate to see that for heavy-shells the semiclassical computation of this overlap vanishes 
\be\label{PsiE} 
\overline{\Psi_i(E)} = 0\,,
\ee 
 since the trajectory of the shell cannot terminate anywhere, and corrections to this value are suppressed in the heavy-shell limit.
 Microscopically, this occurs because $\Psi_i(E)$ does not contain any phase correlation in the erratic entries of the $\mathcal{O}^i$ operator and the signal is washed out semiclassically. 
\begin{figure}[ht]
    \centering
    \includegraphics[width = .35\textwidth]{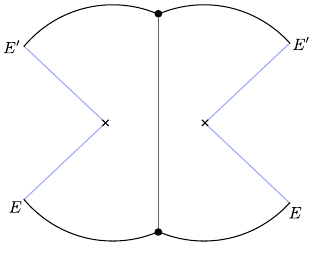}
    \caption{Euclidean pac-man wormhole geometry computing $\overline{\Psi_i(E)\Psi_i(E')^*}$. Note that the solution enforces $E=E'$, as well as $Q_I=Q_I'$. By construction, integrating the on-shell action of the pac-man wormhole over $E$ and the charges $Q_I$ gives the normalization of the PEGS, $Z_{1,i}$. Higher moments such as $\overline{\Psi_i(E)\Psi_j(E')^*\Psi_j(E)\Psi_i(E')^*}$ semiclassically factorize into disconnected copies of the pac-man wormhole geometry. Gluing these copies together to form the semiclassical PEGS wormholes produces additional factors of $e^{-S(E,Q_I)}$ coming from Hayward corner terms.}
    \label{fig:pacman}
\end{figure}

Despite \eqref{PsiE}, we expect that the norms of the wavefunctions $\Psi_i(E)$ computed by the GPI will be non-zero, and that this will follow from non-trivial `microcanonical wormhole' geometries that connect different asymptotic boundaries. Indeed, it is easy to construct the {\it pac-man wormhole}, illustrated in Fig. \ref{fig:pacman}. This geometry provides a saddle point contribution to\footnote{For ease of exposition, and following previous notation, we are omitting left/right labels for the energies and charges, which are independently defined for a given microcanonical window in $\mathcal{H}$. We are also omitting explicitly that the $\delta_{E,E'}$ factors in \eqref{pacman} and \eqref{eq:micropacmansecond} include a $\delta_{Q_I,Q_I'}$ for the charges.}
\be \label{pacman}
\overline{\Psi_i(E)\Psi_j(E')^*} = \delta_{E,E'} \delta_{ij} f_i(E,Q_I) \,.
\ee 
In the heavy-shell limit, the trajectory of the shell pinches off, and the pac-man wormhole yields the on-shell action
\be \label{fEQ}
f_i(E,Q_I) = e^{S(E,Q_I) -{\beta}(E-\mu_I Q_I)}\, Z^{(0)}_i\,,
\ee 
where $S(E,Q_I)$ is the Bekenstein-Hawking entropy of the black hole. The only dependence of the action on the properties of the shell is through the term $Z^{(0)}_i$, which is independent of $E$ and $Q_I$, since the shell is localized in the asymptotic region of the Euclidean solution. This will be a feature of all the moments of $\Psi_i(E)$. It implies that, upon normalization, the shell's contribution $Z^{(0)}_i$ cancels out in the moments of $G^\alpha$. As a consequence, these moments will be solely determined by the black hole parameters. 

In more detail, to obtain the semiclassical matrix $\overline{G_{ij}^\alpha}$  \eqref{microG} we first need the norm of the PEGS, $Z_{1,i}$. Eqs.~\eqref{pacman} and \eqref{fEQ} give it as 
\be 
Z_{1,i} = \int \text{d}E\, \text{d}Q_I\, \overline{\Psi_i(E)\Psi_i(E)^*}\,.
\ee 
From \eqref{eq:normalizationmicro} this implies that $\mathbf{Z}_{1,i}^\alpha = \overline{\Psi_i(E_\alpha)\Psi_i(E_\alpha)^*}$. 

Next, we need the higher moments of $\Psi_i(E)$. All of these moments will be computed by saddle point geometries of the GPI consisting of multiple disconnected copies of the pac-man wormhole sketched in Fig. \ref{fig:pacman}. These copies will contain additional $\delta_{E,E'}$ factors identifying the different energies of each pac-man wormhole.  

We have already seen that the on-shell action of the pac-man wormhole is easy to obtain in the heavy-shell limit. For example, for the second moment, the saddle point contains two copies of the pac-man wormhole in Fig. \ref{fig:pacman}, and the semiclassical contribution is simply
\begin{equation}\label{eq:micropacmansecond}
\overline{\Psi_i(E)\Psi_j(E)^*\Psi_j(E')\Psi_i(E')^*} =  \delta_{E,E'} f_i(E,Q_I)f_j(E,Q_I)\,,
\end{equation}
for $i\neq j$, with $f_i(E,Q_I)$ given again by \eqref{fEQ}. By construction, integrating the on-shell action \eqref{eq:micropacmansecond} over $E$ and $E'$ gives the PEGS wormhole\footnote{Here we are omitting the intrinsic constants $Z_i^{(0)}$ and $Z_i^{(0)}$ since they cancel out in the calculation.}
\be\label{eq:secondmomentintmicro}
Z(2\beta;\mu_I) =\int \text{d}E \text{d}Q_I e^{-S(E,Q_I)} \overline{\Psi_i(E)\Psi_j(E)^*\Psi_j(E)\Psi_i(E)^*}\,.
\ee
The factor of $e^{-S(E,Q_I)}$ arises from removing additional corner terms when gluing the different pac-man wormhole wedges along parts of the bulk fixed-energy boundaries (see \cite{Chua:2023srl} for details on how the gluing of microcanonical wavefunctions is performed).\footnote{In our case there is an additional factor of $2$ in the exponent of the gluing factor in \eqref{eq:micropacmansecond} with respect to eq. (1.9) of \cite{Chua:2023srl} from the fact that the overlap contains two gluings, corresponding to the pair of bra-ket factors. Moreover, there is another factor of $2$ which we leave implicit due to the two-sided nature of our states.}

At the same time, eq. \eqref{eq:micropacmansecond} explains, from the perspective of the GPI, why different microcanonical Gram matrices appear semiclassically uncorrelated, $\overline{G^\alpha_{ij}G^{\alpha'}_{ji}} \propto \delta_{\alpha \alpha'}$. Given \eqref{microG} and \eqref{eq:secondmomentintmicro}, this implies that 
\begin{align} 
\mathbf{Z}_2^\alpha \equiv \sum_{a,a'} \overline{\Psi_i(E_a)\Psi_j(E_a)^*\Psi_j(E_{a'})\Psi_i(E_{a'})^*} \approx\nonumber \\ \approx e^{-S(E_\alpha,Q^\alpha_I)} f_i(E_\alpha,Q^\alpha_I)f_j(E_\alpha,Q^\alpha_I)\,.
\end{align}
Using then \eqref{eq:micropacmansecond}, the microcanonical second moment is
\be 
\overline{G^\alpha_{ij}G^{\alpha'}_{ji}} = \delta_{\alpha \alpha'} \dfrac{\mathbf{Z}_2^\alpha}{\mathbf{Z}_{1,i}^\alpha \mathbf{Z}_{1,j}^\alpha} \approx \delta_{\alpha\alpha'} e^{-S(E_\alpha,Q^\alpha_I)}\,.
\ee

The argument immediately extends to the $n$-th moment of the microcanonical Gram matrix, given that $n$ disconnected copies of the pac-man wormhole of Fig. \ref{fig:pacman} generate the $n$-boundary wormhole for the PEGS, with gluing condition 
\begin{align}
&Z(n\beta;\mu_I) =\nonumber \\  &=\int \text{d}E \text{d}Q_I e^{-(n-1)S(E,Q_I)}\overline{\Psi_{i_1}(E)\Psi_{i_2}(E)^*...\Psi_{i_n}(E)\Psi_{i_1}(E)^*} \,.
\end{align} 
This yields the same universal result as in \eqref{microuniov}.

\subsection{Dimension of the black hole Hilbert space}\label{subsec:dimH}

Having derived in two manners the universal result \eqref{microuniov} for the moments of the overlap matrix between microstates $G_{ij}^\alpha$, we can proceed to steps 2 and 3 of the strategy described in Sec.~\ref{sec:countingstates}. The calculations for the trace of the resolvent have been described earlier \cite{Penington:2019kki} (see also \cite{Hsin:2020mfa,Balasubramanian:2022gmo,Balasubramanian:2022lnw,Boruch:2023trc}), so we shall be brief. 

We take a large family $\mathsf{F}_\Omega$ of heavy-shell microstates, $\Omega \gg 1$, with Gram matrix moments given by \eqref{microuniov}. The expansion \eqref{eq:traceresol} for the semiclassical  $\overline{R^\alpha(\lambda)}$ can be written  as a Schwinger-Dyson equation, which, given the simple form of the moments, can be resummed as
\begin{equation}
\lambda\,\overline{R^\alpha(\lambda)}=\Omega+\,e^S\sum\limits_{n=1}^{\infty}\,\left(\,\dfrac{\overline{R^\alpha(\lambda)}}{e^S}\,\right)^n
=\Omega+\,\frac{e^S\,\overline{R^\alpha(\lambda)}}{e^S-\overline{R^\alpha(\lambda)}}\,.\label{eq:resolvent}
\end{equation}
This is a quadratic equation for $\overline{R^\alpha(\lambda)}$ and it is easily solved.
The discontinuity along the real axis \eqref{eq:densitydisc} gives the density of eigenvalues
\be\label{eq:denG}
\begin{split}  
&\overline{D^\alpha(\lambda)}=\delta(\lambda)\left(\Omega-e^{S}\right)\theta(\Omega-e^{S}) \\[.4cm] 
&+ \frac{e^S}{2\pi\lambda}\sqrt{\,\left[\lambda-\left(1-\Omega^{1/2}\, e^{-S/2} \right)^2\,\right]\left[\,\left(1+\Omega^{1/2} \,e^{-S/2} \right)^2-\lambda \right]}\,.
\end{split}
\ee
The continuous part counts the number of positive eigenvalues, while the singular part counts the number of zero eigenvalues. Applying \eqref{eq:dimspan} we get
\be\label{dimspanfinal}
d_\Omega^\alpha = \text{min}\lbrace \Omega, e^{\Sbh}\rbrace \,,
\ee
which has the same form as \eqref{eq:saturation}. 
Then, the arguments in Sec.~\ref{sec:countingstates} lead us to conclude that the black microstates that we have constructed correctly reproduce the Bekenstein-Hawking entropy
\be 
\log D_\alpha = S=\frac{A}{4G\hbar}\,.
\ee 
With the caveats discussed in Sec.~\ref{subsec:AF}, this result also applies to the asymptotically flat Reissner-Nordström and Kerr black holes.

\section{Corrections}\label{sec:corrections}

We will now see how this framework can incorporate quantum and statistical corrections to the dimension of the black hole Hilbert space.

Quantum corrections from the perspective of the Gibbons-Hawking partition function have been studied in several situations \cite{Maldacena:2016upp, Stanford:2017thb,Ghosh:2019rcj,Iliesiu:2020qvm,Heydeman:2020hhw,Boruch:2022tno,Iliesiu:2022onk,Banerjee:2023quv, Kapec:2023ruw,Rakic:2023vhv,Banerjee:2023gll,Emparan:2023ypa,Banerjee:2010qc,Sen:2012cj,Sen:2012dw}. 
Our aim here is not to obtain any new corrections, but rather to explain how they generally and readily fit in the universal framework of heavy-shell microstates.

To this end, we first recall that the universality relies on the value of the overlaps \eqref{eq:uniwormholen}
which we derived by taking a heavy-shell limit of the saddle point value of the partition functions. 
We will argue that this equation is still valid when we include quantum corrections, e.g., a one-loop determinant, in the path integral computation of the overlap $Z_n^{i_1 \dots i_n}$. In that case, the partition function $Z(\beta; \mu_I)$ in \eqref{eq:uniwormholen} also includes these corrections. This is not immediately obvious, because for $Z(\beta; \mu_I)$ the quantum corrections are added on top of the standard cigar geometry, while for $Z_n^{i_1 \dots i_n}$ they enter on top of the $n$-boundary wormhole geometry with $n$ shells. Nevertheless, we will see that \eqref{eq:uniwormholen} remains valid essentially for the same reason as at tree level: in the heavy-shell limit, the geometries of the kind in Fig.~\ref{fig:PEGSsemiprep} effectively factorize into two copies of a conventional cigar geometry.

With \eqref{eq:uniwormholen} in hand, the arguments in the previous section extend to give $\log D_{\alpha} = S$, where now $S$ is the entropy obtained from the quantum-corrected Gibbons-Hawking partition function.

In Section~\ref{subsec:statcor} we will discuss other corrections to the computation of the number of states, namely those from the statistical nature of the overlaps between gravitational microstates.

\subsection{Quantum corrections near extremality and BPS}

The argument outlined above is made more simply for the quantum corrections near extremality, since we can carry over the discussion of near-extremal PEGS in Section~\ref{crm}, illustrated in Fig.~\ref{fig:twosidedschw}. There, we argued that the throat near the horizon of near-extremal black holes is avoided by heavy shells, and it is in the throat where the leading quantum effects are localized. More concretely, the length of the throat is inversely proportional to the black hole temperature, and the Schwarzian mode that describes the fluctuations in the position of the mouth dominates the low-temperature dynamics (with $T \lesssim G / (r_0^{d-2} L_2^2)$). This Schwarzian mode can be exactly quantized \cite{Maldacena:2016upp,Stanford:2017thb}. In the absence of supersymmetry, this leads to a significant suppression of the density of states at low energies above extremality, while near-BPS throats exhibit a large degeneracy of states at exactly zero-temperature, separated by a gap from the (approximately) continuous finite-temperature spectrum \cite{Ghosh:2019rcj,Iliesiu:2020qvm,Heydeman:2020hhw,Boruch:2022tno}. 

In addition to these quantum $\log T$ effects, there are also more conventional $\log A$ quantum corrections \cite{Banerjee:2010qc,Banerjee:2011jp,Sen:2012kpz,Sen:2012cj}. These come from light fields present in the theory and are thus heavily dependent on the matter content. In a near-extremal black hole, the $\log A$ and $\log T$ terms can be jointly computed without a reduction to the effective two-dimensional theory in the throat  \cite{Iliesiu:2022onk,Rakic:2023vhv}.

Now it becomes clear that in the heavy-shell limit, all this quantum corrected near-horizon dynamics does not affect the properties of the shell far away from the throat. Just like in previous derivations, the shell dependence disappears from the moments of the Gram matrix, and the dominant low-temperature quantum effects are entirely accounted for through the quantum-corrected partition functions that enter in \eqref{eq:uniwormholen}. Each of these receives corrections from a Schwarzian mode around each Euclidean horizon. 

If the black hole is supersymmetric, the shell microstates are initially prepared as non-supersymmetric configurations with finite preparation temperature $\tilde{\beta}^{-1}$. Then they are projected to the supersymmetric ground state by taking the limit $\tilde{\beta}\to\infty$. When working within the supersymmetric JT theory of the throat, BPS microstates of this kind have been constructed in  \cite{Boruch:2023trc} following on \cite{Lin:2022zxd,Lin:2022rzw}. Our construction gives a different family of supersymmetric black hole microstates containing matter behind the horizon, as mentioned above. Again the main difference is illustrated in Fig.~\ref{fig:twosidedschw}. 

The in-throat (super-)JT microstates used in \cite{Boruch:2023trc},  and our out-of-throat universal microstates also differ in the interpretation (or justification) given for the big number of available shell operators. In \cite{Boruch:2023trc}, following up on  \cite{Penington:2019kki}, one \emph{assumes} the existence of a large number of different operators (flavors) at a certain scaling dimension. In our approach, no large number of flavors needs to be invoked. Instead, the higher-dimensional picture allows to build sufficiently different microstates by simply using a single operator inserted a large number of times, in such a way that an arbitrarily large set of shell masses satisfying \eqref{heavyshell} is obtained.

The main advantage of the in-throat microstate constructions is that the simplified dynamics allows to retain a larger degree of control, e.g., over quantum effects, especially with supersymmetry. Out-of-throat heavy-shell microstate constructions, instead, gain in universality. In appendix \ref{app:OverlapsSchwarzian} we extend our arguments  for the in-throat (super-)JT microstates, although the justification of the gaussianity of shell operators is weaker in that case since we cannot properly take a strict heavy-shell limit.

\subsection{Quantum corrections away from extremality}

Black holes far from extremality also receive quantum corrections $\propto \log A$.
Since the computation cannot be localized near the horizon, it is less obvious that the introduction of the shell does not affect the result. In particular, the shell junction imposes non-trivial gluing conditions for the fields that enter the computation of the one-loop determinant.

Nevertheless, we can argue that this does not modify the validity of \eqref{eq:uniwormholen}. Logarithmic corrections are computed via heat kernel methods \cite{Vassilevich:2003xt}, which yield the coefficient of the $\log A$ correction as a local integral of the saddle-point fields, up to contributions from zero modes. In the wormhole geometries, the shell modifies this calculation only through the junction conditions. But these effects have been shown to be entirely localized at the position of the shell \cite{Vassilevich:2003xt,Gilkey:2001mj}.

To illustrate how these modifications vanish in the heavy-shell limit, we can estimate them for a Schwarzschild black hole in four dimensions. Then, $R_{\star} \sim G_N m/2$ at large $m$ and a typical contribution comes from the discontinuity of $K$ across the shell \cite{Gilkey:2001mj},
\be 
\int_{\mathcal{W}} \sqrt{h} [K]^3 \approx - 16 \pi^2 \,.
\ee 
These contributions to the overlaps are therefore universal factors that appear both in the normalization (one-boundary) computation and in the moments of the Gram matrix (several boundaries). They thus cancel in \eqref{eq:OverlapsNonUniversal}, leaving the universal result \eqref{eq:uniwormholen}.

There are also $\log A$ corrections coming from zero modes of linearized fluctuations. To obtain them, \cite{Sen:2012dw} used scaling arguments of the path integral measure, and these would seem unaffected when very heavy shells are included. Our arguments thus seem robust, but since recent one-loop calculations of zero modes have revealed subtleties \cite{Iliesiu:2022onk,Banerjee:2023quv, Rakic:2023vhv}, a more explicit analysis of shells may be useful.

Finally, besides these corrections from infrared quanta, there is another class of quantum (or stringy) corrections due to ultraviolet degrees of freedom, which give rise to higher-dimension operators in the gravitational action. These modify the entropy \eqref{bhe} by terms that are localized on the horizon \cite{Wald:1993nt} and therefore they do not affect the shell contribution to the partition function. In asymptotically AdS spacetimes, the higher-dimension operators also contribute to the effective cosmological constant and to the definition of mass and asymptotic charges. However, once these are accounted for, there are no further effects on the shells and the universal equation for the moments \eqref{eq:uniwormholen} continues to be valid with the corrected partition functions.

\subsection{Statistical corrections}\label{subsec:statcor}

The semiclassical path integral of gravity has only access to phase-correlated properties of the microscopic Gram matrix. This effective description is given in terms of an ensemble of microscopic Gram matrices. In a microscopic theory, the Gram matrix will be a member of such an ensemble. For this reason, there is another class of corrections to the dimension $D_\alpha$, coming from the statistical variances over the ensemble of microscopic Gram matrices. We will estimate them now.

We can compute the variance of $D_\alpha$ from the semiclassical product of resolvent traces
\be 
\overline{R(\lambda)R(\lambda')}|_{\text{conn}} = \sum\limits_{p,q=0}^{\infty}\,\frac{1}{\lambda^{p+1}\lambda^{q+1}}\,\overline{\text{Tr}(G^p)\text{Tr}(G^{q})}|_{\text{conn}}\,.
\ee 

The diagrams contributing to these contractions will be those for which the thin shells propagate between the two sets of boundaries. Microscopically, the shell operators have gaussian random statistics and the contribution of these diagrams to the variance of the rank will vanish identically. Notice this holds in all examples considered in this article. A similar phenomenon was pointed out in \cite{Boruch:2023trc} for matter operators with flavors (but same conformal dimension) coupled to the $\mathcal{N}=2$ super-Schwarzian description of BPS throats. In our case, the large mass limit motivates this effect, given that the microscopic operators become free relative to each other in this limit (see appendix \hyperref[app:FreeProbability]{B}).

In general, off-shell configurations will exist that spoil the exact vanishing of the semiclassical variance of $D_\alpha$ for non-BPS black holes. As argued above, these are not related to the operators that create the microstates but only to spectral variances in the density of states of the Hamiltonian. For near-extremal black holes, these spectral variances can be computed and, when averaged over the microcanonical window, they scale as 
\begin{align}
    \langle (\rho-\rho_0)^2\rangle  \sim e^{-2S_0}(\Delta E_\alpha)^{-2}\;,
\end{align}
from random matrix universality. This gives corrections to the black hole entropy that are suppressed by small factors $\sim 1/S_0$. These errors do not appear for BPS black holes because in those scenarios every microstate has the same energy and, therefore, there is no quantum chaos in the spectrum to start with.

\section{Discussion}\label{sec:discussion}

We have shown that the gravitational path integral provides a universal construction of microstates that account for the entropy of a very wide class of black holes---indeed, as wide as the class for which the Gibbons-Hawking calculus yields a gravitational entropy, possibly including quantum corrections. Although the GPI is the central object both in the Gibbons-Hawking approach and in these microstate constructions, the way that the entropy arises is, as we have emphasized, conceptually different in each of them. But the two approaches are consistent with each other, and this was not a priori obvious: one might have entertained the possibility that, while Gibbons and Hawking found an entropy equal to one-quarter of the area, the microstate construction could have conceivably yielded, say, only one-eighth of it or any other number, indicating that either something was missing or that the GPI was inconsistent. And indeed an inconsistency (essentially Hawking's paradox) would follow had we missed the wormholes: the dimension of the space of microstates would then be infinite. Our main result has been to explain and illustrate how the heavy-shell construction gives a transparent understanding of the universal agreement between these two uses of the GPI.

Can we go beyond the effective statistical description of microstates of the GPI? Having access to the microscopic Gram matrix $G_{ij}$ is likely as hard as solving the holographic CFT at strong coupling. 
The constructions of the kind we have presented suggest that a smooth enough geometry necessarily involves a degree of randomness in the gravitational microstates. Given the chaotic nature of non-extremal black holes, it may not even be sensible to seek a more detailed identification of their microstates of a specific kind. The basis of shell microstates we have employed may not seem the most natural one from the viewpoint of the underlying CFT, but in the gravity picture it does appear natural and simple\footnote{The fuzzball program aims at combining these two viewpoints \cite{Bena:2022rna}.}. As we have argued, it remarkably fulfills the job of furnishing a complete set of Hilbert space states with random statistics, which is universal because it does not rely on any microscopic theory. It is possible that, at least for the states of non-supersymmetric black holes in four or higher dimensions, this is as far as one can go using a bulk geometric picture.

Lower spacetime dimensions may afford a greater degree of control. For instance, in AdS$_3$/CFT$_2$ the conformal and modular symmetries of the CFT, and the isometries of AdS$_3$, determine that the Cardy density of states gives a value of $D_\alpha$ that agrees at leading order with the Bekenstein-Hawking entropy of the BTZ black hole. Our methods recover this result for gravitational theories of AdS$_3$ with massive particles in the spectrum. Could purely gravitational degrees of freedom suffice to account for the entropy?

\section*{Acknowledgments}

We thank Vijay Balasubramanian, Alex Belin, Jan Boruch, Albion Lawrence and Alex Maloney for stimulating conversations. We are grateful to the long-term workshop YITP-T-23-01 held at YITP, Kyoto University, and to the Centro de Ciencias de Benasque Pedro Pascual, where part of this work was done. MS is grateful to the Instituto Balseiro CAB for their hospitality during stages of the project. AC is supported by MICIN through grant PRE2021-098495. AC and RE are supported by AGAUR grant 2017-SGR 754, MICIN grants PID2019-105614GB-C22 and PID2022-136224NB-C22 funded by MCIN/AEI/ 10.13039/501100011033/FEDER, UE, and by the State Research Agency of MICIN through the ‘Unit of
Excellence Maria de Maeztu 2020-2023' award to the Institute of Cosmos Sciences (CEX2019-
000918-M). JM is supported by Conicet, Argentina. MS is supported by the U.S. Department of Energy through DE-SC0009986 and QuantISED DE-SC0020360. AVL is supported by the F.R.S.-FNRS Belgium through conventions FRFC PDRT.1025.14 and IISN 4.4503.15, as well as by funds from the Solvay Family. This preprint has the code BRX-TH-6719.

\appendix
\section{Junction conditions} \label{app:JunctionConditions}

Junction conditions on a generic codimension-1 surface $\mathcal{W}$ have been obtained long ago in the literature \cite{Israel:1966rt,deLaCruz:1967} (we follow the conventions in \cite{Poisson:2009pwt}). Consider two solutions of the Einstein equations (possibly in the presence of gauge fields) which we denote $\mathcal{M}_+$ and $\mathcal{M}_-$ glued along $\mathcal{W}$. We take $n^{\mu}$ to be a unit normal to $\mathcal{W}$ pointing from $\mathcal{M}_-$ to $\mathcal{M}_+$, with $n^{\mu} n_{\mu} = \varepsilon$. For our practical applications, the hypersurface $\mathcal{W}$ will have $\varepsilon = 1$ (both in the Euclidean preparation of states and in their Lorentzian evolution), but we will give the results in general. Let $e_a^{\mu}$ be a basis for the tangent space to $\mathcal{W}$, $h^{\pm}_{ab} = g^{\pm}_{\mu \nu} e_a^{\mu} e_b^{\nu}$ the induced metric from each side, and $K^{\pm}_{ab} = e_a^{\mu} e_b^{\nu} \nabla^{\pm}_{\nu} n_{\mu}$ the extrinsic curvature. The gravitational junction conditions state that
\be\label{eq:junctionGravity}
[h_{ab}] = 0 \,, \quad [K_{ab}] - h_{ab} [K] = - 8 \pi G \epsilon S_{ab} \,,
\ee 
where brackets denote the jump of a given quantity, e.g., $[K_{ab}] = K^+_{ab} - K^-_{ab}$; and $S_{ab}$ is the stress energy tensor in $\mathcal{W}$. More precisely, the full stress energy tensor of the spacetime contains a $\delta$-singular piece localized in $\mathcal{W}$ of the form $T_{\mu \nu} (x) \supset \delta_{\mathcal{W}}(x) S_{\mu \nu} (x)$ with $S^{\mu \nu} = S^{a b} e_a^{\mu} e_b^{\nu}$.

Similar junction conditions can be obtained for the electromagnetic field starting from Maxwell's equations. Decomposing $F_{\mu \nu}$ in $\mathcal{W}$ as $F_{ab} \equiv F_{\mu \nu} e_a^{\mu} e_b^{\nu}$ and $f_a \equiv F_{\mu \nu} e_a^{\mu} n^{\nu}$, we get:
\be \label{eq:junctionMaxwell}
[F_{a b}] = 0 \,, \quad [f_a] = 4 \pi G j_a \,,
\ee
where $j_a$ is the electromagnetic current in $\mathcal{W}$. This means the full current has a component localized in $\mathcal{W}$ of the form $J_{\mu}(x) \supset \delta_{\mathcal{W}}(x) j_{\mu} (x)$ with $j^{\mu} = j^a e_a^{\mu}$.

As an example, we can be more explicit than in the main text about the construction of the Euclidean geometry computing the grand-canonical two-point function $Z_1$ \eqref{eq:norm}. We glue two Reissner-Nordstr\"om solutions with \emph{different} parameters $(M_{\pm}, Q_{\pm})$,
\be
\rd s_{\pm}^2 = f_{\pm}(r_{\pm}) \rd \tau_{\pm}^2 + f_{\pm}^{-1}(r_{\pm}) \rd r_{\pm}^2 + r_{\pm}^2 \rd \Omega_{d-1}^2 \,,
\ee
where
\be
f_{\pm} (r) = r^2 + 1 - \frac{16 \pi G M_{\pm}}{(d-1)V_{\Omega} r^{d-2}} + \frac{32 \pi^2 G^2 Q_{\pm}^2}{(d-1)(d-2) V_{\Omega}^2 r^{2d-4}} \,,
\ee
and the gauge potential is
\be
A_{\pm} = i \left( \frac{4 \pi G Q_{\pm}}{(d-2) V_{\Omega} r_{\pm}^{d-2}} - \Phi \right) \rd \tau_{\pm} \,.
\ee
We glue along a spherically symmetric shell following a curve in the $\tau - r$ plane: $\tau_{\pm} = \mathcal{T}_{\pm} (T)$ and $r_{\pm} = R_{\pm} (T)$. The shell $\mathcal{W}$ also covers homogeneously the $S^{d-1}$, with spherical symmetry reducing the problem to an effective two-dimensional one. The solutions can be glued with the stress energy tensor of a pressureless fluid, $S_{ab} = - \sigma u_a u_b$, and a charge current $j_a = i \sigma_e u_a$ (the sign in $S_{ab}$ and the $i$ in $j_a$ come from Wick rotation to the Euclidean). The vector $u^a = \partial_T^a$ is tangent to curves parametrized by proper length in the $\tau - r$ plane, from which it follows that
\be
\dot{\mathcal{T}}_{\pm} = \mp f^{-1}_{\pm} (R_{\pm}) \sqrt{f_{\pm}(R_{\pm}) - \dot{R}_{\pm}^2} \,.
\ee
The sign is chosen by convention (it just reflects a choice positive $\tau$ orientation, which defines positive and negative charges). At this point, it is already possible to implement the first junction condition in \eqref{eq:junctionGravity}, which imposes $R_+ = R_- \equiv R$ from continuity of the $S^{d-1}$ metric. The radial coordinate is continuous and we will thus suppress $\pm$ indices in it from now on. The normal to $\mathcal{W}$ is
\be
n^{\mu}\partial_{\mu} = \kappa_{\pm} \left( \pm f_{\pm}^{-1} (R) \partial_{\tau_{\pm}} + \sqrt{f_{\pm}(R) - \dot{R}^2} \partial_r \right) ~ .
\ee 
We have included a sign factor $\kappa$ which will be chosen depending on the position of the shell gluing the two black holes. If the shell travels through the exterior of the black hole (so that only the horizon in $\mathcal{M}_-$ is visible in the Euclidean section), we pick $\kappa_- = \kappa_+ = +1$. To build a trapped shell between two horizons as described in the main text and depicted in Fig.~\ref{fig:PEGSsemiprep} we pick $\kappa_- = +1$ and $\kappa_+ = -1$.

We now implement the other gluing conditions. $[F_{ab}] = 0$ is trivially satisfied, and the remaining electromagnetic condition imposes $Q_+ - Q_- = R^{d-1} V_{\Omega} \sigma_e (R) \equiv q$. This is a conservation equation, which can actually be seen to arise from $D_a j^a = 0$ (this must be true whenever there is no electromagnetic current in the bulk solutions to both sides of $\mathcal{W}$). The final gravitational constraint in \eqref{eq:junctionGravity} has two different pieces, one coming from the $T$ direction and the other from the $S^{d-1}$ angles. Together they require $m \equiv R^{d-1} V_{\Omega} \sigma(R)$ to be a constant that we identify with the mass of the shell, and the following evolution equation for $R(T)$:
\be \label{eq:evolutionShell}
\kappa_- \sqrt{f_- - \dot{R}^2} - \kappa_+ \sqrt{f_+ - \dot{R}^2} = \frac{8 \pi G m}{(d-1) V_{\Omega} R^{d-2}} \,.
\ee 
From this equation it follows an effective one-dimensional conservation equation $\dot{R}^2 + V_{\text{eff}} (R) = 0$ with
\begin{align}
\nonumber V_{\text{eff}}(R) = - f_+ (R) + & \left( \frac{M_+ - M_-}{m} - \frac{4 \pi G m}{(d-1)V_{\Omega} R^{d-2}} \right. \\
 & \left. \quad - \frac{2 \pi G (Q_+^2 - Q_-^2)}{m(d-2) V_{\Omega} R^{d-2}} \right)^2  \,.
\end{align}

A solution to the effective one-dimensional problem will not always solve \eqref{eq:evolutionShell}. In order to guarantee this, the sign of the terms within the square roots $f_{\pm}(R) + V_{\text{eff}}(R)$ must be the correct one along the trajectory to obtain the right-hand side. This translates into two inequalities which depend on $\kappa_{\pm}$. For the shell passing between the horizons $\kappa_+ = -1 = - \kappa_-$, we get:
\be
\begin{aligned}
\frac{4 \pi G m^2}{(d-1) V_{\Omega}} + (M_+ - M_-) R^{d-2} - \frac{2 \pi G (Q_+^2 - Q_-^2)}{(d-2) V_{\Omega}} \geq 0 \,, \\
\frac{4 \pi G m^2}{(d-1) V_{\Omega}} - (M_+ - M_-) R^{d-2} + \frac{2 \pi G (Q_+^2 - Q_-^2)}{(d-2) V_{\Omega}} \geq 0 \,. \\
\end{aligned}
\ee
These are generically complicated conditions which must be analyzed on a case by case basis.\footnote{See \cite{Lemos:2021jtm} for an analysis of some cases in $d +1 = 4$ Lorentzian signature.} We can get a flavor of their meaning by looking at a simple situation, e.g., $Q_+ = Q_- = 0$ and $M_+ > M_-$. In that case, the first condition is trivially satisfied, while the second one imposes a \emph{minimum} shell's mass $m^2 \geq (4 \pi G)^{-1} (d-1) V_{\Omega} (M_+ - M_-) R_{\star}^{d-2}$, where $R_{\star}$ is the turning point of the trajectory with $V_{\text{eff}}(R_{\star}) = 0$. As $R$ increases and we approach the boundary, the second condition will be violated, but then we can continue the trajectory bending it over itself by choosing $\kappa_+ = -1$. In the case of gluing the shell in the exterior, $\kappa_+ = \kappa_- = +1$, we revert the second inequality above, getting a \emph{maximum} mass. None of the inequalities is relevant in the $\mathbb{Z}_2$-symmetric configurations studied in the main text, since $Q_+ = Q_-$ and $M_+ = M_-$. But the previous discussion shows that we can glue such solutions along a shell with an arbitrarily large mass $m$, provided it is trapped between two horizons.

\section{Heavy-shell limit and free probability}
\label{app:FreeProbability}

The previous shell operators \eqref{shello} and states \eqref{states} were naturally labeled by the number of operator insertions. This number is dual to the proper mass in the bulk. Since there is no upper bound in the possible number of insertions, it is natural to inquire about the nature of the heavy-shell limit. This is also important given the technical leverage that this limit has given for simplifying calculations and in particular for deriving the key result \eqref{eq:uniwormholen} that universally characterizes the moments of the Gram matrix of overlaps.

It turns out that these results are consistent (indeed equivalent) with a simple interpretation. Consider two shell operators $\mathcal{O}^{m_1}$ and $\mathcal{O}^{m_2}$ with masses $m_1$ and $m_2$. Their effective microscopic description follows from a generalized version \cite{Foini:2018sdb} of the Eigenstate Thermalization Hypothesis \cite{Deutsch:1991msp,Srednicki:1994mfb}, where each operator has its own distribution for its entries in the energy basis. Equivalently, when written in the energy basis, the operators are random matrices with particular distributions. When both masses $m_1$ and $m_2$ are large, and when the difference between them $\vert m_1 -m_2\vert$ is also large, the gravitational description of the shell dynamics implies these shells cannot interact with each other, since the amplitude of interaction decays with the mass difference.\footnote{Showing this in detail is a purely QFT problem, which will be considered elsewhere. Heuristically, the density of states of a QFT in infinite volume is infinite, and the inner product between states with ever increasing mass difference (therefore with ever increasing distance in phase space) will decay to zero. Equivalently, even if the QFT has interactions allowing for such transition, as we increase the mass difference more and more particles need to be created or destroyed, decreasing the probability.} Therefore, the gravitational description of the shell in the heavy-shell limit readily implies that the erratic matrices become independent Gaussian random matrices. It is a simple exercise to verify that from this Gaussian nature (whose variance follows from the norm of the states) one can derive the universal result for the Gram matrix \eqref{eq:uniwormholen}.\footnote{As an example application we consider the $\mathcal{N} = 2$ Schwarzian theory in the next appendix \ref{app:OverlapsSchwarzian}.}

In turn, this Gaussian nature implies that the shell operators with large masses and large mass differences are ``free'' relative to each other, in the sense of free probability. The field of free probability extends the notion of statistical independence for classical variables to the case of non-commutative random variables \cite{Speicher}. Intuitively, two non-commutative random variables $X$ and $Y$ are statistically independent from each other if we can compute the moments of $X+Y$ and those of $XY$ from the moments of $X$ and $Y$ separately. More precisely, one says that a set of non-commutative random variables $X_1,X_2,\dots ,X_s$ are asymptotically free relative to each other when the following holds. Define the ``centered alternating'' random variables 
\be 
C\equiv (X_{i_1}^{m_1}-\langle X_{i_1}^{m_1}\rangle )(X_{i_2}^{m_2}-\langle X_{i_2}^{m_2}\rangle)... (X_{i_r}^{m_r}-\langle X_{i_r}^{m_r}\rangle)\,,
\ee
where $[i_1,... ,i_r]\in [1,..., s]$, $i_{j}\neq i_{j+1}$, and $m_1,\dots ,m_r$ are positive integers. Then the variables are free if the expectation values of all possible $C$'s are zero in the large matrix size limit. In Ref. \cite{Speicher} it is shown how two independent Gaussian random matrices satisfy the previous condition, and are thus mutually free. The crucial input is that for a single Gaussian random matrix of size $N$, the expectation values can be written as
\be \label{expecf}
\langle\textrm{Tr}(X^{2k}_N)\rangle = \sum\limits_{\pi \in P_2 (2k)} N^{-2g_{\pi}}\,,
\ee
where $P_2 (2k)$ is the set of pairings of $2k$ elements and $g_{\pi}$ is the genus of the surface where the pairing $\pi$ can be embedded without crossings. It is remarkable that this follows readily for shells of large masses from the gravitational description, directly classifying possible wormhole geometries in such limit.

Naturally, the leading contribution in \eqref{expecf} comes from the set of non-crossing pairings $NC_2 (2k)$, whose elements $\pi$ have $g_{\pi}=0$. Therefore,  the previous expectation value is the number of non-crossing pairings of $2k$ elements, which is equal to the $k$-th Catalan number, namely
\be \label{TopoETH}
\langle\textrm{Tr}(X^{2k}_N)\rangle \rightarrow \vert NC_2 (2k)\vert = C(k)=\frac{1}{k+1}\binom{2k}{k}\,. 
\ee
This way one arrives at the Wigner semicircle law, appropriate for a Gaussian random matrix, since the Catalan numbers are the moments of this distribution. With this input, a further combinatorial computation allows to prove that two independent Gaussian random matrices are asymptotically free \cite{Speicher}.

It is thus an important problem to show that the dust-shell operators become actually free, relative to each other, in the large mass limit and from a purely microscopic perspective. This would be a microscopic verification of the effective gravitational description. This is a powerful tool because it allows to provide, from a simple combinatorial perspective,  a classification of leading contributions to $n$-boundary saddles, and also of their corrections. More precisely, whenever we have a set of boundary conditions associated with different operator insertions, the average value will be expanded in a topological expansion similar to~\eqref{TopoETH}, where the leading contribution will come from the set of non-crossing pairings and the subleading corrections will organize themselves in a topological expansion controlled by the nature of the remaining pairings.

\section{Overlaps from $\mathcal{N} = 2$ Schwarzian theory}
\label{app:OverlapsSchwarzian}

In this appendix we analyze the case of the $\mathcal{N} = 2$ Schwarzian theory. The partition function and disc correlation functions of the $\mathcal{N}=2$ super JT gravity were computed in \cite{Lin:2022rzw}. These describe statistical aspects of $1/16$ BPS black holes in AdS$_5\times S_5$. More precisely, considering a sector of R-charge $j$, the partition function was found to be
\be 
Z_j\equiv \langle \Psi_{0}^j\vert \Psi_{0}^j\rangle = e^{S_0}\,\cos \pi j\,,
\ee
where $\vert \Psi_{0}^j\rangle$ is the unnormalized vacuum state in the fixed charge sector that is produced via the usual path integral methods. The two point function of an operator of dimension $\Delta$ was found to be
\be 
\textrm{Tr}_{j} (e^{-\beta H}\mathcal{O}_{\Delta}e^{-\beta H}\mathcal{O}_{\Delta})\xrightarrow[\beta\rightarrow\infty]{}Z_j \cos (\pi j) f(\Delta)\,,
\ee
where we have defined the function
\be  
f(\Delta)=\frac{\Delta\,\Gamma(\Delta)^2\,\Gamma (\Delta+\frac{1}{2}\pm j)}{2\pi\,\Gamma (2\Delta)}\;.
\ee
The two point function can be understood as the norm of unnormalized microstates $\vert \Psi_{\Delta}^j\rangle$, created from the vacuum by the application of an operator $\mathcal{O}_{\Delta}$ of dimension $\Delta$ in the Euclidean section of the theory.

This construction is analogous to the PEGS discussed in section \ref{PEGs}, eqs \eqref{eq:generalPEGS} and \eqref{eq:norm}. Indeed, it can be seen as a particular case of our universal construction by thinking of the operators $\mathcal{O}_{\Delta}$ as a dust-shell operator, when constructed in the higher dimensional theory. We just need to translate the scaling dimension to the proper mass of the shell in the usual way \cite{Aharony:1999ti}.

We now follow our strategy to count states. We take a discrete set of states with different $\Delta_i=i\Delta$, with $i=1,...,\Omega $. In the limit of large $\Delta$, the dust-shell operators become relative free, and therefore behave as Gaussian random matrices. From the two point function of the operator we obtain the variance of the Gaussian random matrix, which is $\frac{\cos (\pi j)}{Z_j}f(\Delta)$. Therefore, in the limit of large dimensions and large dimension differences with respect to each other, we can compute the $n$-th moment of the inner product to be
\begin{equation}\label{stsusy}
\overline{\langle \Psi_{\Delta_1}^j\vert \Psi_{\Delta_2}^j\rangle ...  \langle \Psi_{\Delta_n}^j\vert \Psi_{\Delta_1}^j\rangle}=e^{(2-n)S_0}\cos^2 (\pi j)\,\prod\limits_i f(\Delta_i)\,.
\end{equation}
Formally setting all scaling dimensions to be equal, this formula recovers the results of \cite{Boruch:2023trc} concerning the $n$-boundary wormhole.

This perspective adds two aspects. The first one is that we do not need to assume the existence of a large number of operators at a certain scaling dimension $\Delta$. Assuming this is equivalent to the west coast model \cite{Penington:2019kki}, where one assumes an infinite set of flavors that effectively describe a large global symmetry in the semiclassical theory, leading in particular to $\overline{\langle \Psi_{\Delta_i}\vert \Psi_{\Delta_j}\rangle}=\delta_{ij}$. The difficulty then arises in translating these flavors at a fixed scaling dimension $\Delta$ into actual degrees of freedom in top-down models. This would require, for instance, assuming that the microcanonical degeneracy of the holographic CFT close to a given $\Delta$ is as large as wanted, to accommodate so many flavours. This is presumably not the case, given that the CFT has finite entropy, which is actually the original problem. This conundrum is evaded when one chooses states created by operators with different scaling dimensions, leading to \eqref{stsusy}. The operators in question can be further chosen to be dust-shell operators, so changing the scaling dimension is just accomplished by changing the number of operator insertions of a certain seed light conformal primary. Therefore, in our construction, we only need to assume the presence of a low-energy field, which can be typically identified in several ways in top-down models. 

The second aspect is that, from the present perspective, it seems clear that the gravitational computation described in \cite{Boruch:2023trc} implicitly assumes the gaussianity of the inserted operators. This assumption is generically unjustified. In typical top-down scenarios, operator insertions will have nontrivial interactions between them. Again, this is physically accounted for by choosing dust-shell operators and taking the number of operator insertions to be large, so that the non-gaussianities become subleading. Although there might be other ways to achieve gaussianity (or relative asymptotic freedom), the heavy-shell limit discussed is particularly convenient.

With the statistics \eqref{stsusy} of the inner products, we can compute the Hilbert space dimension in the usual manner. Normalizing the states we obtain
\be  
\frac{\overline{\langle \Psi_{\Delta_1}^j\vert \Psi_{\Delta_2}^j\rangle ...  \langle \Psi_{\Delta_n}^j\vert \Psi_{\Delta_1}^j\rangle}}{\prod\limits_{i}\overline{\langle \Psi_{\Delta_i}^j\vert \Psi_{\Delta_i}^j\rangle }}=(e^{S_0}\cos (\pi j))^{2(1-n)}\,.
\ee
Inserting this in the resolvent equation\footnote{The dependence on the different scaling dimensions disappears in the resolvent equation because of the cancellations between numerator and denominator in the previous equation. This follows from the factorization of the $n$-th point functions as two point functions. It goes in line with the general perspective in the present article, supporting the universality of the construction. Notice that the resolvent equation analyzed in \cite{Boruch:2023trc} depended on $\Delta$ just because the states considered were unnormalized. Normalizing the states before computing the Hilbert space dimension gives rise to the usual resolvent equation.} \eqref{eq:traceresol} we conclude that the Hilbert space dimension spanned by this precise set of $\Omega$ microstates is
\be
\text{dim}(\text{span}\lbrace F_\Omega\rbrace) = \text{min}\lbrace \Omega, Z_j\rbrace \,,
\ee
which provides a derivation of the entropy of these black holes.

\bibliographystyle{utphys}
\bibliography{main}

\end{document}